\documentclass[longauth]{aa}

\usepackage{graphicx}
\usepackage{txfonts}

\usepackage{hyperref}
\usepackage{multirow}
\usepackage{multicol}
\usepackage{textcomp}
\usepackage{xcolor}
\usepackage{soul}
\begin{document}

\title{A search for accreting young companions embedded in circumstellar disks: \\High-contrast H$\alpha$ imaging with VLT/SPHERE\thanks{Based on observations collected at the Paranal Observatory, ESO (Chile). Program ID: 096.C-0248(B), 096.C-0267(A),096.C-0267(B), 095.C-0273(A), 095.C-0298(A)}\fnmsep\thanks{Figures 1, 6 and D.1 are only available in electronic form at the CDS via anonymous ftp to \url{cdsarc.u-strasbg.fr (130.79.128.5)} or via \url{http://cdsweb.u-strasbg.fr/cgi-bin/qcat?J/A+A/}}}

\subtitle{}

\author{G. Cugno\inst{\ref{instch1}}
\and S.~P. Quanz\inst{\ref{instch1},\ref{nccr}}
\and S. Hunziker\inst{\ref{instch1}}
\and T. Stolker\inst{\ref{instch1}}
\and H.~M. Schmid\inst{\ref{instch1}}
\and H.~Avenhaus\inst{\ref{instd1}}
\and P.~Baudoz\inst{\ref{instf4}}
\and A.~J. Bohn\inst{\ref{instnl3}}
\and M.~Bonnefoy\inst{\ref{instf1}}
\and E.~Buenzli\inst{\ref{instch1}}
\and G.~Chauvin\inst{\ref{instf1},\ref{instcl1}} 
\and A.~Cheetham \inst{\ref{instch2}}
\and S.~Desidera\inst{\ref{insti1}}
\and C.~Dominik\inst{\ref{instnl2}}
\and P.~Feautrier\inst{\ref{instf1}}
\and M.~Feldt\inst{\ref{instd1}}      
\and C.~Ginski\inst{\ref{instnl3}}  
\and J.~H.~Girard\inst{\ref{instusa2}, \ref{instf1}}      
\and R.~Gratton\inst{\ref{insti1}}
\and J.~Hagelberg\inst{\ref{instch1}}
\and E.~Hugot\inst{\ref{instf3}}
\and M.~Janson\inst{\ref{instsw1}}
\and A.-M.~Lagrange\inst{\ref{instf1}}
\and M.~Langlois\inst{\ref{instf3},\ref{instf7}}  
\and Y.~Magnard\inst{\ref{instf1}}
\and A.-L.~Maire \inst{\ref{instd1}}
\and F.~Menard\inst{\ref{instf1},\ref{instf2}} 
\and M.~Meyer\inst{\ref{instusa1},\ref{instch1}}
\and J.~Milli\inst{\ref{insteso2}}
\and C.~Mordasini\inst{\ref{instch3}}
\and C.~Pinte\inst{\ref{instau1}, \ref{instf1}}
\and J.~Pragt\inst{\ref{instnl1}}
\and R.~Roelfsema\inst{\ref{instnl1}}
\and F.~Rigal\inst{\ref{instnl1}}  
\and J.~Szul\'agyi\inst{\ref{instch4}}
\and R.~van Boekel\inst{\ref{instd1}}
\and G.~van der Plas\inst{\ref{instf1}}
\and A.~Vigan\inst{\ref{instf3}}
\and Z.~Wahhaj\inst{\ref{insteso2}}
\and A.~Zurlo\inst{\ref{instf3}, \ref{instcl2}, \ref{instcl3}}
}

\institute{ETH Zurich, Institute for Particle Physics and Astrophysics, Wolfgang-Pauli-Strasse 27, CH-8093 Zurich, Switzerland\label{instch1}
\and National Center of Competence in Research "PlanetS"(\url{http://nccr-planets.ch})\label{nccr} 
\and Max-Planck-Institut f\"{u}r Astronomie, K\"{o}nigstuhl 17, 69117 Heidelberg, Germany\label{instd1}
\and LESIA, CNRS, Observatoire de Paris, Universit\'{e} Paris Diderot, UPMC, 5 place J. Janssen, 92190 Meudon, France\label{instf4}
\and Leiden Observatory, Leiden University, P.O. Box 9513, 2300 RA Leiden, The Netherlands\label{instnl3}
\and Univ. Grenoble Alpes, CNRS, IPAG, F-38000 Grenoble, France\label{instf1}
\and Unidad Mixta International Franco-Chilena de Astronomia, CNRS/INSU UMI 3386 and Departemento de Astronomia, Universidad de Chile, Casilla 36-D, Santiago, Chile\label{instcl1}
\and Geneva Observatory, University of Geneva, Chemin des Mailettes 51, 1290 Versoix, Switzerland\label{instch2}
\and INAF - Osservatorio Astronomico di Padova, Vicolo dell'Osservatorio 5, 35122 Padova, Italy\label{insti1}
\and Anton Pannekoek Astronomical Institute, University of Amsterdam, PO Box 94249, 1090 GE Amsterdam, The Netherlands\label{instnl2}
\and Space Telescope Science Institute, Baltimore 21218, MD, USA\label{instusa2}
\and Aix Marseille Universit\'{e}, CNRS, CNES, LAM, Marseille, France\label{instf3}
\and Department of Astronomy, Stockholm University, AlbaNova University Center, 106 91 Stockholm, Sweden\label{instsw1}
\and Centre de Recherche Astrophysique de Lyon, CNRS/ENSL Universit\'{e} Lyon 1, 9 av. Ch. Andr\'{e}, 69561 Saint-Genis-Laval, France\label{instf7}
\and CNRS, IPAG, 38000 Grenoble, France\label{instf2}
\and The University of Michigan, Ann Arbor, Mi 48109, USA\label{instusa1}
\and European Southern Observatory, Alonso de Cordova 3107, Casilla 19001 Vitacura, Santiago 19, Chile\label{insteso2}
\and Physikalisches Institut, Universit\"at Bern, Gesellschaftsstrasse 6, 3012 Bern, Switzerland\label{instch3}
\and Monash Centre for Astrophysics (MoCa) and School of Physics and Astronomy, Monash University, Clayton Vic 3800, Australia\label{instau1}
\and NOVA Optical Infrared Instrumentation Group at ASTRON, Oude Hoogeveensedijk 4, 7991 PD Dwingeloo, The Netherlands\label{instnl1}
\and Institute for Computational Science, University of Zurich, Winterthurerstrasse 190, 8057 Zurich, Switzerland\label{instch4}
\and N\'ucleo de Astronom\'ia, Facultad de Ingenier\'ia y Ciencias, Universidad Diego Portales, Av. Ejercito 441, Santiago, Chile\label{instcl2}
\and Escuela de Ingenier\'ia Industrial, Facultad de Ingenier\'ia y Ciencias, Universidad Diego Portales, Av. Ejercito 441, Santiago, Chile\label{instcl3}
\\\\\email{gabriele.cugno@phys.ethz.ch}
}

%

\date{Received --- ; accepted --- }

\abstract{}{}{}{}{}

\abstract
{In recent years, our understanding of giant planet formation progressed substantially. There have even been detections of a few young protoplanet candidates still embedded in the circumstellar disks of their host stars. The exact physics that describes the accretion of material from the circumstellar disk onto the suspected circumplanetary disk and eventually onto the young, forming planet is still an open question.}
{We seek to detect and quantify observables related to accretion processes occurring locally in circumstellar disks, which could be attributed to young forming planets. We focus on objects known to host protoplanet candidates and/or disk structures thought to be the result of interactions with planets. }
{We analyzed observations of six young stars (age $3.5-10$ Myr) and their surrounding environments with the SPHERE/ZIMPOL instrument on the Very Large Telescope (VLT) in the H$\alpha$ filter (656 nm) and a nearby continuum filter (644.9 nm). 
We applied several point spread function (PSF) subtraction techniques to reach the highest possible contrast near the primary star, specifically investigating regions where forming companions were claimed or have been suggested based on observed disk morphology.
}
{We redetect the known accreting M-star companion HD142527 B with the highest published signal to noise to date in both H$\alpha$ and the continuum. We derive new astrometry  ($r = 62.8^{+2.1}_{-2.7}$ mas and $\text{PA} = (98.7\,\pm1.8)^\circ$) and photometry ($\Delta$N\_Ha=$6.3^{+0.2}_{-0.3}$ mag, $\Delta$B\_Ha=$6.7\pm0.2$ mag and $\Delta$Cnt\_Ha=$7.3^{+0.3}_{-0.2}$ mag) 
for the companion in agreement with previous studies, and estimate its mass accretion rate ($\dot{M}\approx1-2\,\times10^{-10}\,M_\odot\text{ yr}^{-1}$). A faint point-like source around HD135344 B (SAO206462) is also investigated, but a second deeper observation is required to reveal its nature. No other companions are detected. In the framework of our assumptions we estimate detection limits at the locations of companion candidates around HD100546, HD169142, and MWC\,758 and calculate that processes involving H$\alpha$ fluxes larger than $\sim8\times10^{-14}-10^{-15}\,\text{erg/s/cm}^2$ ($\dot{M}>10^{-10}-10^{-12}\,M_\odot\text{ yr}^{-1}$) can be excluded. Furthermore, flux upper limits of $\sim10^{-14}-10^{-15}\,\text{erg/s/cm}^2$ ($\dot{M}<10^{-11}-10^{-12}\,M_\odot \text{ yr}^{-1}$) are estimated within the gaps identified in the disks surrounding HD135344\,B and TW Hya. The derived luminosity limits exclude H$\alpha$ signatures at levels similar to those previously detected for the accreting planet candidate LkCa15\,b.}
{}


\titlerunning{A search for accreting young companions embedded in circumstellar disks} 
\maketitle


\section{Introduction}
Providing an empirical basis for gas giant planet formation models and theories requires the detection of young objects in their natal environment, i.e., when they are still embedded in the gas and dust-rich circumstellar disk surrounding their host star. The primary scientific goals of studying planet formation are as follows: To understand where gas giant planet formation takes place, for example, at what separations from the host star and under which physical and chemical conditions in the disk; how formation occurs, i.e., via the classical core accretion process \citep{Pollack1996}  or a modified version of that process \citep[e.g., pebble accretion,][]{Lambrechts2012} or direct gravitational collapse \citep{Boss1997}); and the properties of the suspected circumplanetary disks (CPDs).\\
While in recent years high-contrast, high spatial resolution imaging observations of circumstellar disks have revealed an impressive diversity in circumstellar disk structure and morphology, the number of directly detected planet candidates embedded in those disks is still small \citep[LkCa15\,b, HD100546\,b, HD169142\,b, MWC\,758\,b, PDS\,70\,b;][]{KrausIreland2012,Quanz2013_discovery,reggiani2014,biller2014,Reggiani2017, Keppler2018}. To identify these objects, high-contrast exoplanet imaging can be used. These observations are typically performed at near- to mid-infrared wavelengths using an adaptive optics-assisted high-resolution camera. In addition to the intrinsic luminosity of the still contracting young gas giant planet, the surrounding CPD, if treated as a classical accretion disk, contributes significantly to fluxes beyond 3$\,\mu$m wavelength \citep{zhu2015, Eisner2015}, potentially easing the detection of young forming gas giants at these wavelengths. While the majority of the forming planet candidates mentioned above were detected in this way, it has also been realized that the signature from a circumstellar disk itself can sometimes mimic that of a point source after PSF subtraction and image post-processing \citep[e.g.,][]{Follette2017, ligi2017}. As a consequence, it is possible that some of the aforementioned candidates are false positives.

Another approach is to look for direct signatures of the suspected CPDs, such as their dust continuum emission or their kinematic imprint in high-resolution molecular line data \citep{Perez2015,Szulagyi2018}. In one case, spectro-astrometry using CO line emission was used to constrain the existence and orbit of a young planet candidate \citep{Brittain2013, Brittain2014}. Moreover, \cite{Pinte2018} and \cite{Teague2018} suggested the presence of embedded planets orbiting HD163296 from local deviations from Keplerian rotation in the protoplanetary disk.
A further indirect way to infer the existence of a young, forming planet is to search for localized differences in the gas chemistry of the circumstellar disk, as the planet provides extra energy to the chemical network in its vicinity \citep{Cleeves2015}.

Finally, it is possible to look for accretion signatures from gas falling onto the planet and its CPD. Accretion shocks are able to excite or ionize the hydrogen atoms, which then radiate recombination emission lines, such as H$\alpha$, when returning to lower energy states \citep[e.g.,][]{calvetgullbring1998, Szulagyi2017, Marleau2017}.
High-contrast imaging using H$\alpha$ filters was already successfully applied in three cases. 
Using angular spectral differential  imaging (ASDI) with the Magellan Adaptive Optics System (MagAO), \cite{close2014} detected H$\alpha$ excess emission from the M-star companion orbiting the Herbig Ae/Be star HD142527, and \cite{sallum2015} also used MagAO to identify at least one accreting companion candidate located in the gap of the transition disk around LkCa15. The accretion signature was found at a position very similar to the predicted orbital position of one of the faint point sources detected by \cite{KrausIreland2012}, attributed to a forming planetary system. Most recently, \cite{Wagner2018} have claimed the detection of H$\alpha$ emission from the young planet PDS70\,b using MagAO, albeit with comparatively low statistical significance (3.9$\sigma$). 

In this paper we present a set of H$\alpha$ high-contrast imaging data for six young stars, aiming at the detection of potential accretion signatures from the (suspected) young planets embedded in the circumstellar disks of the stars. 
The paper is structured as follows: In Section \ref{sec:sample} we discuss the observations and target stars. We explain the data reduction in Section \ref{sec:data_reduction} and present our analyses in Section \ref{sec:analysis}. In Section \ref{sec:discussion} we discuss our results in a broader context and conclude in Section \ref{sec:conclusions}.


\section{Observations and target sample}
\label{sec:sample}
\subsection{Observations}
\begin{table*}[h!]
\caption{\label{tab_obs} Summary of observations.}
\centering
\begin{tabular}{llllllllll}
\hline\hline\noalign{\smallskip}
Object                  & H$\alpha$     & Obs. date     &  Prog. ID             &  DIT\tablefootmark{b}  & \# of  & Field                         & Mean    & $\tau_0$\tablefootmark{c}  & Mean   \\
                        &       Filter\tablefootmark{a}         &[dd.mm.yyyy] &                               & [s]   &       DITs    & rotation [$^\circ$]  &                airmass  & [ms]   & seeing\tablefootmark{d} [as]      \\\hline
\noalign{\smallskip}
\multirow{2}{*}{HD142527}  & B\_Ha & 31.03.2016 & 096.C-0248(B)         & 30 & 70 & 47.8  & 1.06  &$2.7\pm0.2$ & $0.71\pm0.06$ \\ \cline{2-10} 
\noalign{\smallskip}
                                          & N\_Ha   & 31.03.2016   &    096.C-0248(B)           & 30   & 70 & 48.6 & 1.05  &$2.7\pm0.3$  & $0.69\pm0.07$ \\\hline
\noalign{\smallskip}                     
                  HD135344 B        &N\_Ha & 31.03.2016 & 096.C-0248(B)          & 50 & 107  & 71.7  & 1.04  &$4.4\pm1.2$  & $0.47\pm0.17$   \\ \hline
 \noalign{\smallskip}                    
                  TW Hya        &B\_Ha & 23.03.2016 & 096.C-0267(B)     & 80 & 131 & 134.1  & 1.16 &$1.4\pm0.4$  & $1.33\pm0.53$   \\ \hline
\noalign{\smallskip}                     
                  HD100546          &B\_Ha & 23.04.2015 & 095.C-0273(A)          & 10 & 1104\tablefootmark{e} & 68.3\tablefootmark{e}  & 1.46 &$1.7\pm0.2$  & $0.98\pm0.28$ \\ \hline
\noalign{\smallskip}                     
                  HD169142         & B\_Ha & 09.05.2015  & 095.C-0298(A)  &   50    &     90      &   123.2    &  1.01     &$1.4\pm0.1$   & $1.24\pm0.04$  \\ \hline
 \noalign{\smallskip}                    
                  MWC\,758        &B\_Ha & 30.12.2015 & 096.C-0267(A)   & 60 & 194 & 54.8  & 1.63  &$3.2\pm0.8$  & $1.39\pm0.24$  \\ 
\noalign{\smallskip}\hline\hline
\end{tabular}
\tablefoot{\tablefoottext{a}{Each dataset consists of data obtained in one of the two H$\alpha$ filters and simultaneous data taken with the continuum filter inserted in the other ZIMPOL camera.}\tablefoottext{b}{DIT = Detector integration time, i.e., exposure time per image frame.}\tablefoottext{c}{Coherence time}.\tablefoottext{d}{Mean DIMM seeing measured during the observation.}\tablefoottext{e}{As we explain in Section~\ref{sec:Analysis_HD100546} and Appendix~\ref{App_3}, for this dataset a frame selection was applied, which reduced the number of frames to 366 and the field rotation to $20.7^\circ$.}
}
\end{table*}
The data were all obtained with the ZIMPOL sub-instrument of the adaptive optics (AO) assisted high-contrast imager SPHERE \citep{Beuzit2008, Petit2008, Fusco2016}, which is installed at the Very Large Telescope (VLT) of the European Southern Observatory (ESO) on Paranal in Chile. A detailed description of ZIMPOL can be found in \cite{Schmid2018}. Some of the data were collected within the context of the Guaranteed Time Observations (GTO) program of the SPHERE consortium; others were obtained in other programs and downloaded from the ESO data archive (program IDs are listed in Table \ref{tab_obs}). We focused on objects that are known from other observations to host forming planet candidates that still need to be confirmed (HD100546, HD169142, and MWC\,758)\footnote{In the discussion (Section \ref{sec:discussion}) we also include the analysis of a dataset of LkCa15 (PI: Huelamo) to set our results in context, but the data were poor in quality and hence not included in the main part of the paper.}, objects known to host accreting stellar companions (HD142527), and objects that have well-studied circumstellar disks with spatially resolved substructures (gaps, cavities, or spiral arms), possibly suggesting planet formation activities (HD135344 B and TW Hya). 
All data were taken in the noncoronagraphic imaging mode of ZIMPOL 
using an H$\alpha$ filter in one camera arm and a nearby continuum filter simultaneously in the other
arm (Cont\_Ha; $\lambda_c=644.9$ nm, $\Delta\lambda=3.83$ nm). As the data were observed in different programs, we sometimes used the narrow H$\alpha$ filter (N\_Ha; $\lambda_c=656.53$ nm, $\Delta\lambda=0.75$ nm) and sometimes the broad H$\alpha$ filter (B\_Ha; $\lambda_c=655.6$ nm, $\Delta\lambda=5.35$ nm). A more complete description of these filters can be found in \cite{schmid2017}. To establish which filter allows for the highest contrast performance, we used HD142527 and its accreting companion \citep{close2014} as a test target and switched between the N\_Ha and the B\_Ha filter every ten frames within the same observing sequence. All datasets were observed in pupil-stabilized mode to enable angular differential imaging \citep[ADI;][]{marois2006}. The fundamental properties of the target stars are given in Table~\ref{tab_stars}, while a summary of the datasets is given in Table~\ref{tab_obs}. \\
We note that because of the intrinsic properties of the polarization beam splitter used by ZIMPOL, polarized light might preferentially end up in one of the two arms, causing a systematic uncertainty in the relative photometry between the continuum and H$\alpha$ frames. The inclined mirrors in the telescope and the instrument introduce di-attenuation (e.g., higher reflectivity for $I_\perp$ than $I_\parallel$) and polarization cross talks, so that the transmissions in imaging mode to the $I_\perp$ and $I_\parallel$ arm depend on the telescope pointing direction. This effect is at the level of a few percent (about $\pm 5~\%$), but unfortunately the dependence on the instrument configuration has not been determined yet. 
We discuss its potential impact on our analyses in Appendix \ref{App:Beamsplitter}, even though we did not take this effect into account since it is small and could not be precisely quantified. 

\subsection{Target sample}
\textit{HD142527}\\\\
HD142527 is known to have a prominent circumstellar disk \citep[e.g.,][]{fukagawa2006, Canovas2013, Avenhaus2014_HD142527} and a close-in M star companion \citep[HD142527 B;][]{biller2012,rodigas2014,lacour2016,Christiaens2018, Claudi2018} that shows signatures of ongoing accretion in H$\alpha$ emission \citep{close2014}. This companion orbits in a large, optically thin cavity within the circumstellar disk stretching from $\sim0\farcs07$ to $\sim1\farcs0$ \citep[e.g.,][]{Fukagawa2013, Avenhaus2014_HD142527}, and it is likely that this companion is at least partially responsible for clearing the gap by accretion of disk material \citep{biller2012, Price2018}. \cite{Avenhaus2017} obtained polarimetric differential imaging data with SPHERE/ZIMPOL in the very broad band  \cite[VBB, as defined in ][]{Schmid2018} optical filter, revealing new substructures, and resolving the innermost regions of the disk (down to $0\farcs025$). In addition, extended polarized emission was detected at the position of HD142527 B, possibly due to dust in a circumsecondary disk. \cite{Christiaens2018} extracted a medium-resolution spectrum of the companion and suggested a mass of $0.34\pm0.06\,M_\odot$. This value is a factor of $\sim3$ larger than that estimated by spectral energy distribution (SED) fitting \citep[][$M=0.13\pm0.03\,M_\odot$]{lacour2016}. Thanks to the accreting close-in companion, this system is the ideal target to optimize the H$\alpha$ observing strategy with SPHERE/ZIMPOL and also the data reduction. \\\\
\textit{HD135344 B}\\\\
HD135344 B (SAO206462) is surrounded by a transition disk that was spatially resolved at various wavelengths. Continuum (sub-)millimeter images presented by \cite{Andrews2011} and \cite{vanderMarel2016} revealed a disk cavity with an outer radius of $0\farcs32$. In polarimetric differential imaging (PDI) observations in the near-infrared (NIR), the outer radius of the cavity appears to be at $0\farcs18$, and the difference in apparent size was interpreted as a potential indication for a companion orbiting in the cavity \citep{garufi2013}. Data obtained in PDI mode also revealed two prominent, symmetric spiral arms \citep{muto2012,garufi2013, stolker2016}.
\cite{Vicente2011} and \cite{maire2017} searched for planets in the system using NIR NACO and SPHERE high-contrast imaging data, but did not find any. Using hot start evolutionary models these authors derived upper limits for the mass of potential giant planets around HD135344 B (3 M$_J$ beyond $0\farcs7$).\\
\\
\textit{TW Hya}\\\\
TW Hya is the nearest T Tauri star to Earth. Its almost face-on transitional disk  \citep[$i\sim7\pm1^\circ$;][]{qi2004} shows multiple rings and gaps in both dust continuum and scattered light data. Hubble Space Telescope (HST) scattered light images from \cite{Debes2013} first allowed the identification of a gap at $\sim1\farcs48$. Later, \cite{Akiyama2015} observed in $H$-band polarized images a gap at a separation of $\sim0\farcs41$. Using Atacama Large Millimeter Array (ALMA), \cite{Andrews2016} identified gaps from the radial profile of the 870 $\mu$m continuum emission at $0\farcs41$, $0\farcs68$ and $0\farcs80$.
Finally, \cite{vanboekel2017} obtained SPHERE images in PDI and ADI modes at optical and NIR wavelengths, 
and identified three gaps at $0\farcs11$, $0\farcs39,$ and $1\farcs57$ from the central star. A clear gap was also identified by \cite{Rapson2015} at a separation of $0\farcs43$ in Gemini/GPI polarimetric images and the largest gap at $r\simeq1\farcs52$ has also been observed in CO emission with ALMA \citep{Huang2018}.
\\
\begin{table*}[t!]
\caption{\label{tab_stars} Target sample.}
\centering
\begin{tabular}{lllllll}
\hline\hline\noalign{\smallskip}
Object                  & RA                                    & DEC           & Spec. type      & $m_R$ [mag] & Distance [pc] & Age [Myr]\\\hline
\noalign{\smallskip}
HD142527         & 15$^h$56$^m$41.89$^s$    & -42$^\circ$19$'$23\farcs27 & F6III & 7.91 &$157.3\pm1.2$ & $8.1^{+1.9}_{-1.6}$ \\
\noalign{\smallskip}
HD135344 B          &  15$^h$15$^m$48.44$^s$    &  -37$^\circ$09$'$16\farcs03           & F8V & 8.45 &     $135.9\pm1.4$& $9\pm2$ \\
\noalign{\smallskip}
TW Hya          &  11$^h$01$^m$51.90$^s$        &  -34$^\circ$42$'$17\farcs03           & K6Ve & $10.43\pm0.1$ &   $60.1\pm0.1$ & $\sim10$ \\
\noalign{\smallskip}
HD100546              &  11$^h$33$^m$25.44$^s$ & -70$^\circ$11$'$41\farcs24& B9Vne & 8.78 & $110.0\pm0.6$ & $7\pm1.5$ \\
\noalign{\smallskip}
HD169142          &  18$^h$24$^m$29.78$^s$      &  -29$^\circ$46$'$49\farcs32           & B9V &   8.0 & $114.0\pm0.8$ & $\sim6$ \\
\noalign{\smallskip}
MWC\,758          &  05$^h$30$^m$27.53$^s$      &  -25$^\circ$19$'$57\farcs08           & A8Ve & $9.20\pm0.01$ &   $160.3\pm1.7$ & $3.5\pm2$ \\
\noalign{\smallskip}
\noalign{\smallskip}\hline\hline
\end{tabular}
\tablefoot{Coordinates and spectral types are taken from SIMBAD, R-magnitudes are taken from the NOMAD catalog \citep{Zacharias2004} for HD142527 and HD169142, from the APASS catalog \citep{Henden2016} for HD135344\,B, and from the UCAC4 catalog \citep{Zacharias2012} for the other targets. Distances are from GAIA data release 2 \citep{Gaia2018}. The ages -- from top to bottom -- are taken from \cite{Fairlamb2015}, \cite{mueller2011}, \cite{Weinberger2013}, \cite{Fairlamb2015}, \cite{Grady2007}, and \cite{Meeus2012}.}\\

\end{table*}

\textit{HD100546}\\\\
The disk around HD100546 was also spatially resolved in scattered light and dust continuum emission in different bands \citep[e.g.,][]{Augereau2001,Quanz2011,Avenhaus2014,Walsh2014, Pineda2014}. The disk appears to be almost, but not completely, devoid of dusty material at radii between a few and 13 AU. This gap could be due to the interaction with a young forming planet, and \cite{Brittain2013,Brittain2014} suggested the presence of a companion orbiting the star at $0\farcs13$, based on high-resolution NIR spectro-astrometry of CO emission lines. Another protoplanet candidate was claimed by \cite{Quanz2013_discovery} using $L'$ band high-contrast imaging data. The object was found at $0\farcs48\pm0\farcs04$ from the central star, at a position angle (PA) of $(8.9\,\pm0.9)^\circ$, with an apparent magnitude of $L'$=$13.2\pm0.4$ mag. \cite{quanz2015} reobserved HD100546 in different bands ($L'$, $M'$, $K_s$) and detected the object again in the first two filters. 
Based on the colors and observed morphology these authors suggested that the data are best explained by a forming planet surrounded by a circumplanetary disk. Later, \cite{Currie2015} recovered HD100546 b from $H$-band integral field spectroscopy (IFS) with the Gemini Planet Imager \citep[GPI;][]{Macintosh2006} and identified a second putative point source c closer to the star ($r_\mathrm{proj}\sim0\farcs14$) potentially related to the candidate identified by \citet{Brittain2013,Brittain2014}. More recently, \cite{Rameau2017} demonstrated that the emission related to HD100546 b appears to be stationary and its spectrum is inconsistent with any type of low temperature objects. Furthermore, they obtained H$\alpha$ images with the MagAO instrument to search for accretion signatures, but no point source was detected at either the b or  c position, and they placed upper limits on the accretion luminosity ($L_\mathrm{acc}<1.7\times10^{-4}\;L_\odot$). The same data were analyzed by \cite{Follette2017}, together with other H$\alpha$ images (MagAO), $H$ band spectra (GPI), and Y band polarimetric images (GPI). Their data exclude that HD100546\,c is emitting in H$\alpha$ with $L_{H\alpha}>1.57\times10^{-4} L_\odot$.\\
\\
\textit{HD169142}\\\\
HD169142 is surrounded by a nearly face-on pre-transitional disk. 
Using PDI images, \cite{quanz2013} found an unresolved disk rim at $0\farcs17$ and an annular gap between $0\farcs28$ and $0\farcs49$. These results were confirmed by \cite{osorio2014}, who investigated the thermal emission ($\lambda=7$ mm) of large dust grains in the HD169142 disk, identifying two annular cavities ($\sim0\farcs16-0\farcs21$ and $\sim0\farcs28-0\farcs48$). The latter authors also identified a point source candidate in the middle of the outer cavity at a distance of $0\farcs34$ and PA $\sim175^\circ$. 
\cite{biller2014} and \cite{reggiani2014} observed a point-like feature in NaCo $L'$ data at the outer edge of the inner cavity (separation = $0\farcs11-0\farcs16$ and PA=$0^\circ-7.4^\circ$).  
Observations in other bands ($H$, $K_S$, $z_p$) with the Magellan Clay Telescope (MagAO/MCT) and with GPI in the $J$ band  failed to confirm the detection \citep{biller2014,reggiani2014}, but revealed another candidate point source albeit with low signal-to-noise ratio \citep[S/N;][]{biller2014}.
In a recent paper, \cite{ligi2017} explained the latter \cite{biller2014} detection with a bright spot in the ring of scattered light from the disk rim, potentially following Keplerian motion. 
\cite{Pohl2017} and \cite{Bertrang2018} compared different disk and dust evolutionary models to SPHERE $J$-band and VBB PDI observations. Both works tried to reproduce and explain the complex morphological structures observed in the disk and conclude that planet-disk interaction is occurring in the system, even though there is no clearly confirmed protoplanet identified to date.\\
\\
\textit{MWC\,758}\\\\
MWC\,758 is surrounded by a pre-transitional disk \citep[e.g.,][]{Grady2013}. \cite{Andrews2011} found an inner cavity of $\sim$55 AU based on dust continuum observations, which was, however, not observed in scattered light \citep{Grady2013, Benisty2015}. Nevertheless, PDI and direct imaging from the latter studies revealed two large spiral arms. A third spiral arm has been suggested based on VLT/NaCo $L'$ data by \cite{Reggiani2017}, together with the claim of the detection of a point-like source embedded in the disk at $(111\pm4)$ mas. This object was observed in two separate datasets from 2015 and 2016 at comparable separations from the star, but different PAs, which was possibly due to orbital motion. The contrast of this object relative to the central star in the $L'$ band is $\sim7$ mag, which, according to the BT-Settl atmospheric models \citep{Allard2012}, corresponds to the photospheric emission of a 41-64 $M_J$ object for the age of the star. 
More recently, ALMA observations from \cite{Boehler2018} traced the large dust continuum emission from the disk. Two rings at $0\farcs37$ and $0\farcs53$ were discovered that are probably related to two clumps with large surface density of millimeter dust and a large cavity of $\sim0\farcs26$ in radius. Finally, \cite{Huelamo2018} observed MWC 758 in H$\alpha$ with SPHERE/ZIMPOL, reaching an upper limit for the line luminosity of $L_\mathrm{H_\alpha}\lesssim5\times10^{-5}L_\odot$ (corresponding to a contrast of 7.6 mag) at the separation of the protoplanet candidate. No other point-like features were detected.\\

\section{Data reduction}
\label{sec:data_reduction}
The basic data reduction steps were carried out with the ZIMPOL pipeline developed and maintained at ETH Z\"urich. The pipeline remapped the original 7.2 mas / pixel $\times$ 3.6 mas / pixel onto a square grid with an effective pixel scale of 3.6 mas / pixel $\times$ 3.6 mas / pixel (1024 $\times$ 1024 pixels). Afterward, the bias was subtracted and a flat-field correction was applied. We then aligned the individual images by fitting a Moffat profile to the stellar point spread functions (PSFs) and shifting the images using bilinear interpolation. The pipeline also calculated the parallactic angle for each individual frame and added the information to the image header. Finally, we split up the image stacks into individual frames and grouped them together according to their filter, resulting in two image stacks for each object: one for an H$\alpha$ filter and one for the continuum filter\footnote{For HD142527 we have four image stacks as we used both the N\_Ha and the B\_Ha filter during the observing sequence.}. In general, all images were included in the analysis if not specifically mentioned in the individual subsections. The images in these stacks were cropped to a size of $1\farcs08\,\times1\,\farcs08$ centered on the star. This allowed us to focus our PSF subtraction efforts on the contrast dominated regime of the images. The removal of the stellar PSF was performed in three different ways: ADI, spectral differential imaging (SDI), and ASDI (a two-step combination of SDI and ADI).

To perform ADI, we fed the stacks into our {\tt PynPoint} pipeline \citep{amaraquanz2012,amara2015, Stolker2018}. The {\tt PynPoint} package uses principal component analysis (PCA) to model and subtract the stellar PSF in all individual images before they are derotated to a common field orientation and mean-combined. To investigate the impact on the final contrast performance for all
objects, we varied the number of principal components (PCs) used to fit the stellar PSF and the size of the inner mask that is used to cover the central core of the stellar PSF prior to the PCA. No frame selection based on the field rotation was applied, meaning that all the images were considered for the analysis, regardless of the difference in parallactic angle. 
The SDI approach aims at reducing the stellar PSF using the fact that all features arising from the parent star (such as Airy pattern and speckles) scale spatially with wavelength $\lambda$, while the position of a physical object on the detector is independent of $\lambda$. The underlying assumption is that, given that $\lambda_c$ is similar in all filters, the continuum flux density is the same at all wavelengths. To this end, modified versions of the continuum images were created. First, they were multiplied with the ratio of the effective filter widths to normalize the throughput of the continuum filter relative to the H$\alpha$ filter\footnote{This approach ignores any potential color effects between the filters, which, given their narrow band widths, should, however, not cause any significant systematic offsets.}. Then, they were spatially stretched using spline interpolation in radial direction, going out from the image center, by the ratio of the central wavelengths of the filters to align the speckle patterns. Because of the possibly different SED shapes of our objects with respect to the standard calibration star used in \cite{schmid2017} to determine the central wavelengths $\lambda_c$ of the filters, it is possible that $\lambda_c$ is slightly shifted for each object. This effect, however, is expected to alter the upscaling factor by at most 0.4\% for B\_Ha (assuming the unrealistic case in which $\lambda_c$ is at the edge of the filter), which is the broadest filter we used. This is negligible at very small separations from the star, where speckles dominate the noise. Values for filter central wavelengths and filter equivalent widths can be found in Table 5 of \cite{schmid2017}. The modified continuum images were then subtracted from the images taken simultaneously with the H$\alpha$ filter, leaving only H$\alpha$ line flux emitted from the primary star and potential companions. As a final step, the images resulting from the subtraction are derotated to a common field orientation and mean-combined. It is worth noting that if, as a result of the stretching, a potential point-source emitting a significant amount of continuum flux moves by more than $\lambda/D$, the signal strength in the H$\alpha$ image is only marginally changed in the SDI subtraction step, and only the speckle noise is reduced. If this is not the case, this subtraction step yields a significant reduction of the source signal in addition to the reduction of the
speckle noise. For SPHERE/ZIMPOL H$\alpha$ imaging, a conservative SDI subtraction without substantial signal removal is achieved for angular separations $\gtrsim0\farcs90$ ($\sim250$ pixels). Nevertheless, this technique is expected to enhance the S/N of accreting planetary companions even at smaller separations, since young planets are not expected to emit a considerable amount of optical radiation in the continuum. In this case, the absence of a continuum signal guarantees that the image subtraction leaves the H$\alpha$ signal of the companion unchanged and only reduces the speckle residuals. Therefore, for this science case, there is no penalty for using SDI.\\  
To perform ASDI, the SDI (H$\alpha$-Cnt\_H$\alpha$) subtracted images are fed into the PCA pipeline to subtract any remaining residuals. During the analysis we varied the same parameters as described  for simple ADI. 
The HD142527 dataset was used to compare the different sensitivities achieved when applying ADI, SDI, and ASDI. The results are discussed in Section \ref{sec:setup_performance} and Appendix \ref{App_1}.\\
With ZIMPOL in imaging mode, there is a constant offset of $(135.99\pm0.11)^\circ$ between the parallactic angle and the PA of the camera in sky coordinates \citep{Maire2016}. A preliminary astrometric calibration showed, however, that this reference frame has to be rotated by $(-2.0\pm 0.5)^\circ$ to align images with north pointing to the top (Ginski et al., in preparation). This means that overall, for every PSF subtraction technique, the final images have to be rotated by $(134\pm0.5)^\circ$ in the counterclockwise direction. 

\begin{figure*}[t!]
\centering
\includegraphics[width=\hsize]{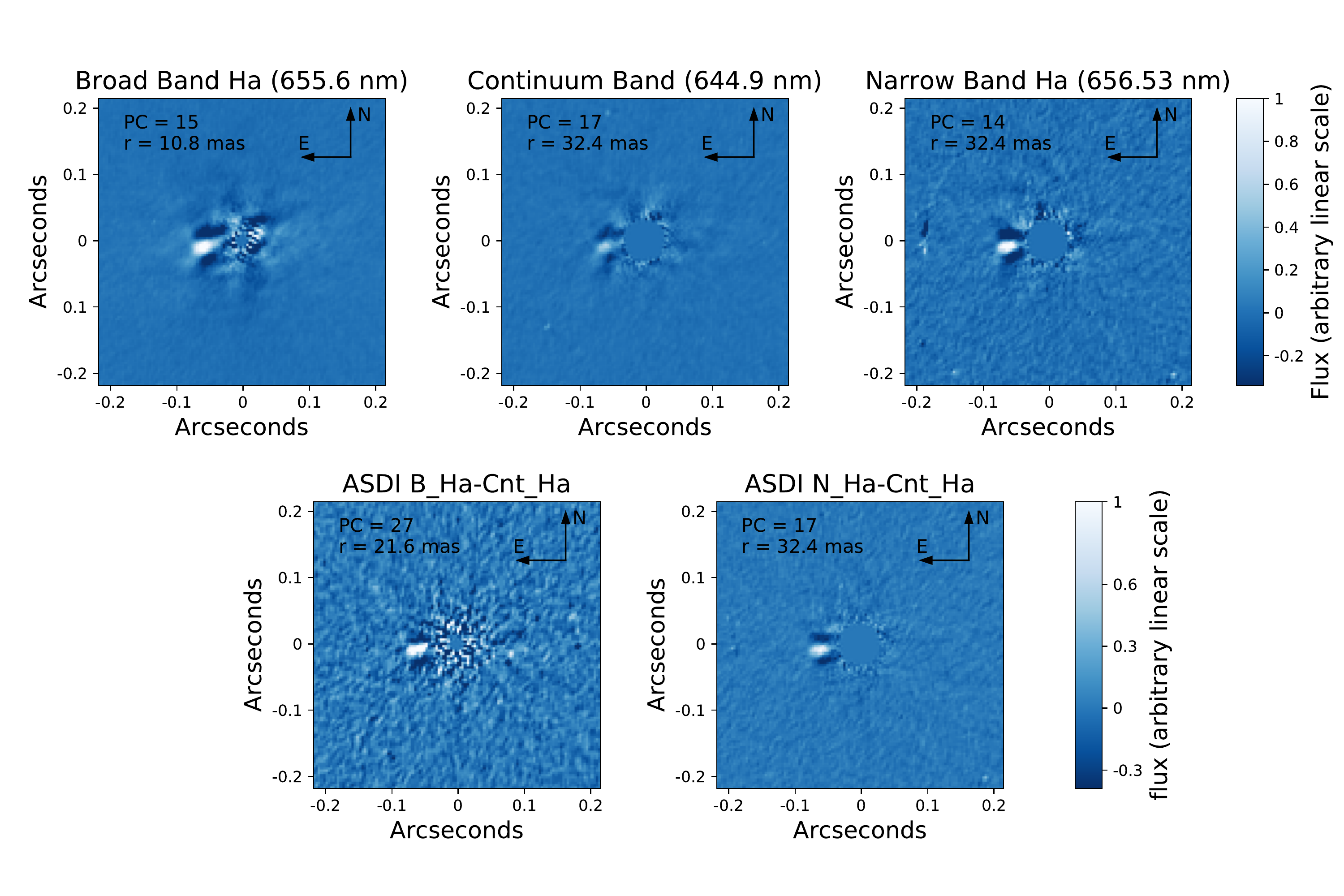}
\caption{Final ADI and ASDI reduced images of HD142527. \textit{Top row:} B\_Ha, Cnt\_Ha, and N\_Ha filter images resulting in the lowest FPFs ($1.5\times10^{-11}$, $2.2\times10^{-9}$, and $<10^{-17}$, corresponding to S/Ns of 13.1, 9.8, and 26.6, respectively). \textit{Bottom row:} final images after ASDI reduction for B\_Ha-Cnt\_Ha and N\_Ha-Cnt\_Ha frames ($4.4\times10^{-16}$ and $<10^{-17}$, corresponding to S/Ns of 22.7 and 27.6). We give the number of subtracted PCs and the radius of the central mask in milliarcseconds in the top left corner of each image. The color scales are different for the two rows. Because all images of the top row have the same color stretch, the detection appears weaker in the continuum band.}
\label{hd142527b}
\end{figure*}

\begin{figure}[h!]
\centering
\includegraphics[width=\hsize]{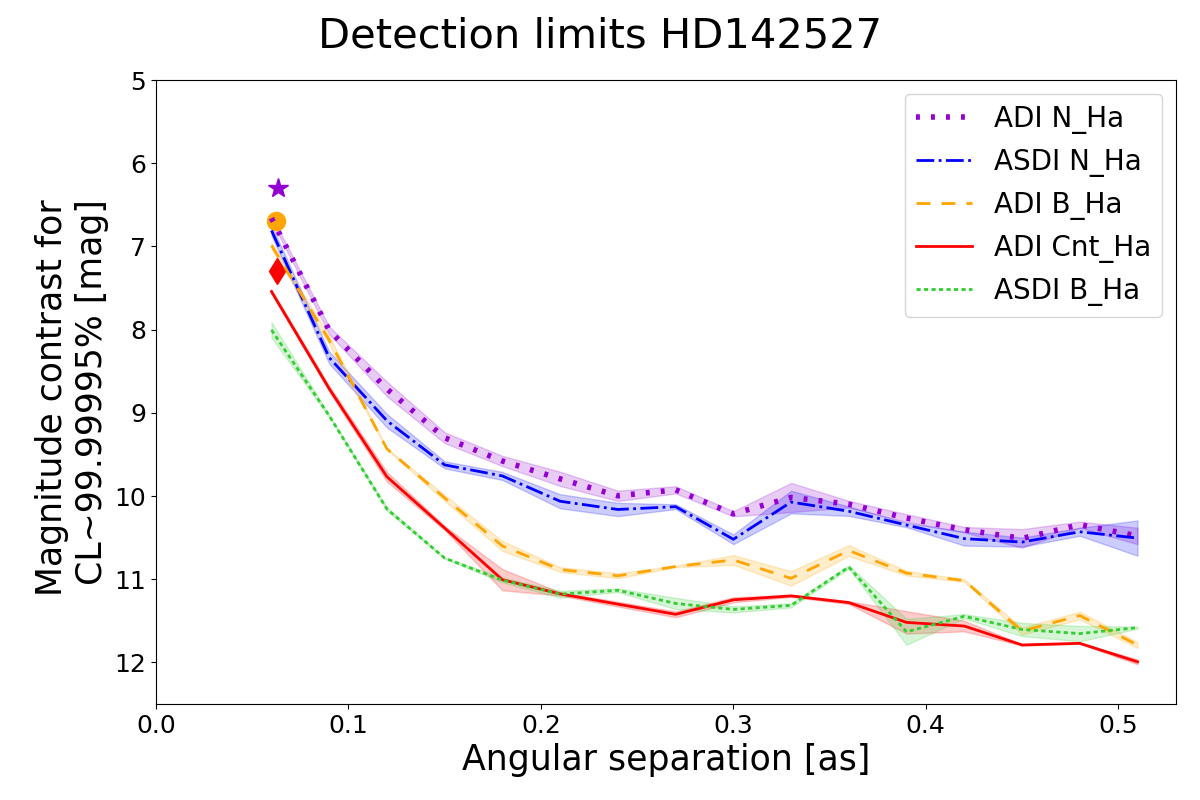}
\caption{Contrast curves for HD142527. The colored shaded regions around each curve represent the standard deviation of the achieved contrast at the 6 azimuthal positions considered at each separation. The markers (red diamond, orange circle, and violet star) represent the contrast of HD142527 B.}
\label{contrast_hd142527}
\end{figure}

\section{Analysis and results}
\label{sec:analysis}
\subsection{HD142527 B: The accreting M-star companion}
\subsubsection{Comparing the performance of multiple observational setups}
\label{sec:setup_performance}
In this section, we quantitatively compare the detection performance for multiple filter combinations and PSF subtraction techniques and establish the best strategy for future high-contrast H$\alpha$ observations with SPHERE/ZIMPOL. For the analysis, the HD142527 dataset was used; during the data reduction, no further frame selection was applied. The final images of HD142527 clearly show the presence of the M-star companion east of the central star. The signal is detected in all filters with ADI (B\_Ha, N\_Ha, and Cnt\_Ha) and ASDI (in both continuum-subtracted B\_Ha and N\_Ha images) over a broad range of PCs and also for different image and inner mask sizes (see Figure~\ref{hd142527b}). \\
We used the prescription from \cite{mawet2014} to compute the false positive fraction (FPF) as a metric to quantify the confidence in the detection. The flux is measured in apertures of diameter $\lambda /D$ (16.5 mas) at the position of the signal and in equally spaced reference apertures placed at the same separation but with different PAs, so that there is no overlap between these angles and the remaining azimuthal space is filled. These apertures sample the noise at the separation of the companion. Since the apertures closest to the signal are dominated by negative wings from the PSF subtraction process, they were ignored. Then, we used Equation 9 and Equation 10 from \cite{mawet2014} to calculate S/N and FPF from these apertures. This calculation takes into account the small number of apertures that sample the noise and uses the Student t-distribution to calculate the confidence of a detection. The wider wings of the t-distribution enable a better match to a non-Gaussian residual speckle noise than the normal distribution. 
However, the true FPF values could be higher if the wings of the true noise distribution are higher than those of the t-distribution\footnote{As an example, Figure 7 of \cite{mawet2014} shows how the t-distribution produces lower FPF values than the case where speckle noise follows more closely a modified Rician distribution. Nevertheless, it has been shown that applying ADI removes the correlated component of the noise leaving quasi-Gaussian residuals \citep{Marois2008}.}. \\
The narrow N\_Ha filter delivers a significantly lower FPF than the broader B\_Ha filter over a wide range of PCs (see Figure~\ref{fpf_HD142527} in Appendix \ref{App_1}). Figure~\ref{fpf_HD142527} also shows that the combination of SDI and ADI yields lower FPF values than only ADI for both filters. Applying ASDI on N\_Ha images is hence the preferred choice for future high-contrast imaging programs with SPHERE/ZIMPOL in the speckle-limited regime close to the star. Furthermore, as shown in Figure \ref{fig:obs_param} and explained in Appendix~\ref{App_2}, it is crucial to plan observations maximizing the field rotation to best modulate and subtract the stellar PSF and to achieve higher sensitivities. 

In Figure~\ref{contrast_hd142527} we show the resulting contrast curves for the three filters for a confidence level (CL) of 99.99995\%. For each dataset (B\_Ha, N\_Ha, and Cnt\_Ha) and technique (ADI and ASDI), we calculated the contrast curves for different numbers of PCs (between 10 and 30 in steps of 5) after removing the companion (see Section \ref{sec:characterization}). From each set of curves, we only considered the best achievable contrast at each separation from the central star. The presence of H$\alpha$ line emission from the central star made SDI an inefficient technique to search for faint objects at small angular separations.

To derive the contrast curves, artificial companions with varying contrast were inserted at six different PAs (separated by 60$^\circ$) and in steps of $0\farcs03$ in the radial direction. As the stellar PSF was unsaturated in all individual frames, the artificial companions were obtained by shifting and flux-scaling the stellar PSFs and then adding these companions to the original frames. Also for the calculation of the ASDI contrast curves, the original H$\alpha$ filter images, containing underlying continuum and H$\alpha$ line emission, were used to create artificial secondary signals. For each reduction run only one artificial companion was inserted at a time to keep the PCs as similar as possible to the original reduction. The brightness of the artificial signals was reduced/increased until their FPF corresponded to a detection with a CL of 99.99995\% (i.e., a FPF of 2.5$\times10^{-7}$), corresponding to $\approx$5$\sigma$ whether Gaussian noise was assumed. An inner mask with a radius of $0\farcs02$ was used to exclude the central parts dominated by the stellar signal. The colored shaded regions around each curve represent the standard deviation of the contrast achieved at that specific separation within the six PAs.

It is important to note that, while in Figure~\ref{fpf_HD142527} the N\_Ha filter provides the lowest FPF for the companion, Figure~\ref{contrast_hd142527} seems to suggest that the B\_Ha filter provides a better contrast performance. However, this is an effect from the way the contrast analysis is performed. As described above, the stellar PSF was used as a template for the artificial planets, as it is usually done in high-contrast imaging data analysis. The flux distribution within a given filter can vary significantly depending on the object. In this specific case, HD142527 B is known to have H$\alpha$ excess emission, hence the flux within either H$\alpha$ filter is strongly dominated by line emission ($\sim$50\% in B\_Ha and $\sim$83\% in N\_Ha filter) and a contribution from the optical continuum can be neglected. The primary shows, however, strong and non-negligible optical continuum emission that contributes to the flux observed in the H$\alpha$ filters. Indeed, for the primary, only 10\% and 56\% of the flux in the B\_Ha and N\_Ha filters are attributable to line emission. Hence, when using the stellar PSF as  template for artificial planets, we obtain a better contrast performance for the B\_Ha filter as it contains overall more flux. In reality, however, if the goal is to detect H$\alpha$ line emission from low-mass accreting companions, the N\_Ha filter is to be preferred. Finally, as found by \cite{sallum2015} for the planet candidate LkCa15 b, the fact that ASDI curves reach a deeper contrast confirms that this technique, in particular close to the star, is more effective and should be preferred to search for H$\alpha$ accretion signals. 

\subsubsection{Quantifying the H$\alpha$ detection}
\label{sec:characterization}
The clear detection of the M-star companion in our images allows us to determine its contrast in all the filters and its position relative to the primary at the epoch of observation. For this purpose, we applied the Hessian matrix approach \citep{quanz2015} and calculated the sum of the absolute values of the determinants of Hessian matrices in the vicinity of the companion's signal. The Hessian matrix represents the second derivative of an n-dimensional function and its determinant is a measure for the curvature of the surface described by the function. This method allows for a simultaneous determination of the position and the flux contrast of the companion and we applied a Nelder-Mead \citep{NelderMead1965} simplex algorithm to minimize the curvature, i.e., the determinants of the Hessian matrices. We inserted negative, flux-rescaled stellar PSFs at different locations and with varying brightness in the input images and computed the resulting curvature within a region of interest (ROI) around the companion after PSF subtraction\footnote{For this analysis we used an image size of $0\farcs36\times0\farcs36$  to speed up the computation and an inner mask of 10.8 mas (radius).}. To reduce pixel-to-pixel variations after the PSF-subtraction step and allow for a more robust determination of the curvature, we convolved the images with a Gaussian kernel with a full width at half maximum (FWHM) of 8.3 mas ($\approx0.35$ of the FWHM of the stellar PSF, which was calculated to be 23.7 mas on average). To fully include the companion's signal, the ROI was chosen to be $(43.2\times43.2)$ mas around the peak flux detected in the original set of PSF subtracted images. Within the ROI, the determinants of the Hessian matrices in 10,000 evenly spaced positions on a fixed grid (every 0.43 mas) were calculated and summed up.

\begin{figure}[t!]
\centering
\includegraphics[width=7cm]{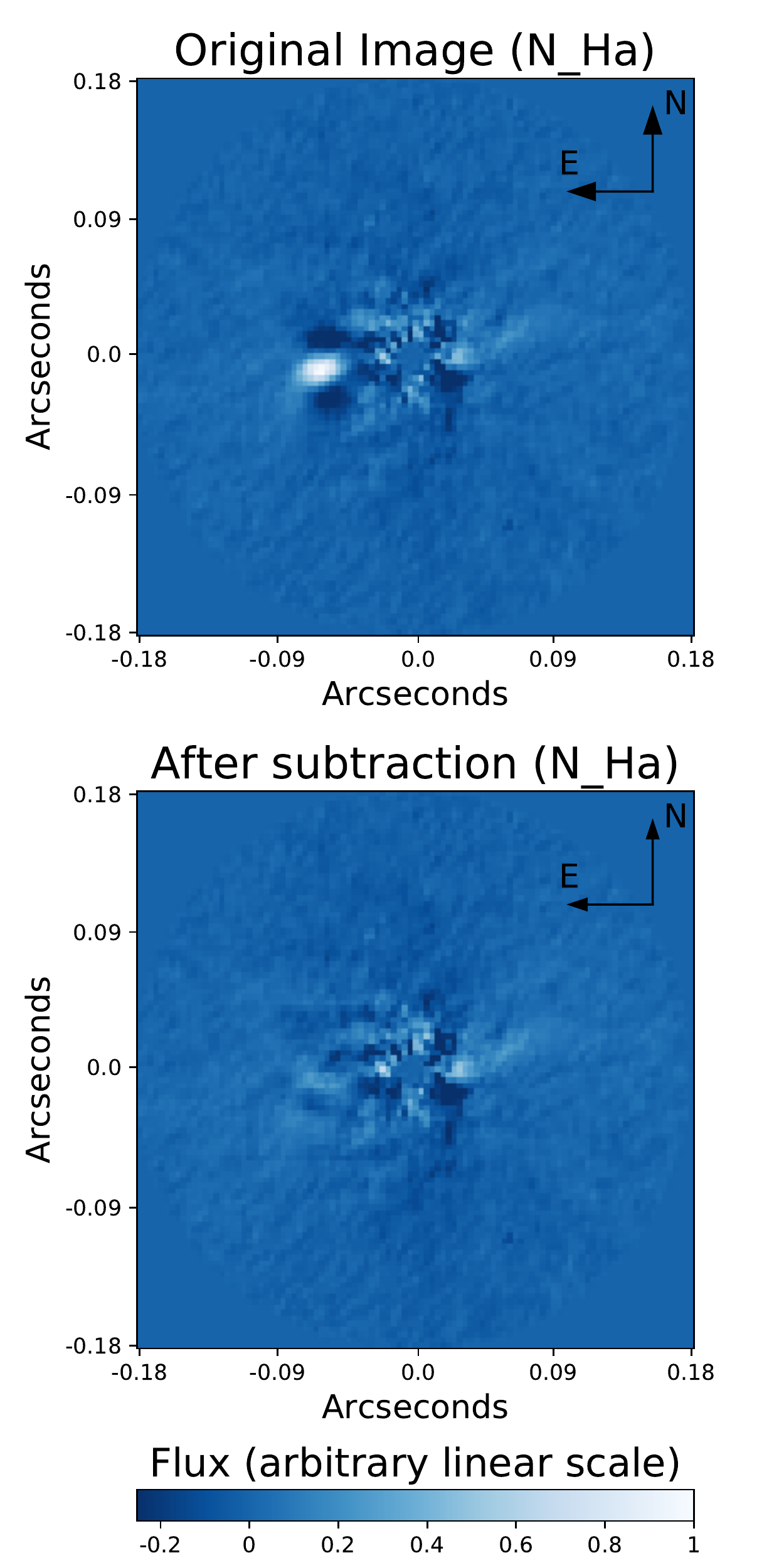}
\caption{Image of HD142527 before (top panel) and after (bottom panel) the insertion of the negative companion resulting from the Hessian matrix algorithm. The image flux scale is the same in both images. In this case 14 PCs were subtracted and a mask of 10.8 mas (radius) was applied on the $101\times$ 101 pixels images of the N\_Ha stack.}
\label{fig:Hesse_result}
\end{figure}

For the optimization algorithm to converge, we need to provide a threshold criterion: if the change in the parameters (position and contrast) between two consecutive iterations is less than a given tolerance, the algorithm has converged and the optimization returns those values for contrast and position. The absolute tolerance for the convergence was set to be 0.1\footnote{This is an absolute value, meaning that if the sum of the determinants can be lowered only using steps in pixels and contrast lower than 0.1, then the algorithm stops.}, as this      value is the precision to which artificial signals can be inserted into the image grid. This value applies for all the investigated parameters (position and contrast). Errors in the separation and PA measurements take into account the tolerance given for the converging algorithm and the finite grid. Errors in the contrast magnitude only consider the uncertainty due to the tolerance of the optimization.
To account for systematic uncertainties in the companion's location and contrast resulting from varying self-subtraction effects in reductions with different numbers of PCs, we ran the Hessian matrix algorithm for reductions with PCs in the range between 13 and 29 and considered the average of each parameter as final result. This range of PCs corresponds to FPF values below $2.5\times10^{-7}$ (see Figure~\ref{fpf_HD142527}). 
To quantify the overall uncertainties in separation, PA, and contrast in a conservative way, we considered the maximum/minimum value (including measurement errors) among the set of results for the specific parameter and computed its difference from the mean.
In Figure \ref{fig:Hesse_result}, we present the results from this approach for the N\_Ha dataset and show the comparison between the original residual image and the image with the companion successfully removed.

\begin{table*}[h!]
\caption{\label{tab:Ha_star} Summary of the stellar fluxes measured in the different filters in our ZIMPOL data and the derived H$\alpha$ line fluxes for our targets (last column). The extinction values $A_{H\alpha}$ were estimated as described in Section \ref{sec:photometry_HD142527} from $A_V$.}
\centering
\begin{tabular}{llllll}
\hline\hline\noalign{\smallskip}
Object                  & $A_V$ [mag] &$A_{H\alpha}$ [mag] & $F^*_{\text{F\_H}\alpha}$ [erg/s/cm$^2$] & $F^*_{\text{Cnt\_H}\alpha}$ [erg/s/cm$^2$] & $F^*_{H\alpha}$   [erg/s/cm$^2$]\\\hline
\noalign{\smallskip}
HD142527 (N\_Ha)  & <0.05\tablefootmark{a} & 0.04 & $3.0\pm0.8\times10^{-11}$ & $6.1\pm0.2\times10^{-11}$ & $1.7\pm0.8\times10^{-11}$  \\ \hline 
\noalign{\smallskip}
HD142527 (B\_Ha)  & <0.05\tablefootmark{a} &0.04 & $9.7\pm0.8\times10^{-11}$ & $6.1\pm0.2\times10^{-11}$ & $1.0\pm0.5\times10^{-11}$  \\ \hline 
\noalign{\smallskip}
HD142527 B (N\_Ha)  & <0.05\tablefootmark{a} & 0.04 & $9.1^{+3.5}_{-2.9}\times10^{-14}$ & $7.4^{+1.4}_{-2.1}\times10^{-14}$ & $7.6^{+3.5}_{-2.9}\times10^{-14}$  \\ \noalign{\smallskip}\hline 
\noalign{\smallskip}
HD142527 B (B\_Ha)  & <0.05\tablefootmark{a} &0.04 & $2.0\pm0.4\times10^{-13}$ & $7.4^{+1.4}_{-2.1}\times10^{-14}$ & $1.0^{+0.5}_{-0.4}\times10^{-13}$  \\ \noalign{\smallskip}\hline 
\noalign{\smallskip}
HD135344 B  & 0.23 \tablefootmark{a} &0.19 & $3.1\pm1.0\times10^{-11}$ & $4.9\pm0.6\times10^{-11}$ & $1.8\pm0.8\times10^{-11}$ \\ \hline 
\noalign{\smallskip}
TW Hya  & 0.0\tablefootmark{b} & 0.0 & $9.9\pm0.4\times10^{-11}$ & $1.5\pm0.05\times10^{-11}$ &$7.8\pm0.3\times10^{-11}$  \\ \hline 
\noalign{\smallskip}
HD100546  &<0.05\tablefootmark{a} & 0.04 & $4.2\pm0.2\times10^{-10}$ & $1.6\pm0.1\times10^{-10}$ & $1.7\pm0.2\times10^{-10}$  \\ \hline 
\noalign{\smallskip}
HD169142  & 0.43\tablefootmark{c} & 0.35 & $1.1\pm0.1\times10^{-10}$& $7.4\pm0.2\times10^{-11}$ & $3.2\pm4.4\times10^{-12}$  \\ \hline 
\noalign{\smallskip}
MWC758  & 0.22\tablefootmark{d} & 0.18 & $8.1\pm0.7\times10^{-11}$ & $5.3\pm0.2\times10^{-11}$ &$6.3\pm3.7\times10^{-12}$  \\ 
\noalign{\smallskip}\hline\hline
\end{tabular}
\tablebib{\tablefoottext{a}{\cite{Fairlamb2015}.}\tablefoottext{b}{\cite{uyama2017}.}\tablefoottext{c}{\cite{Fedele2017}.}\tablefoottext{d}{\cite{vandenAncker1998}.}}
\end{table*}

\subsubsection{Astrometry}
\label{sec:HD142527_astrometry}
The previously described algorithm was used to determine the best combination of separation, PA, and magnitude contrast for HD142527 B. In the N\_Ha data the companion is located at $63.3^{+1.3}_{-1.0}$ mas from the primary star, in the B\_Ha dataset at $62.3^{+1.7}_{-2.2}$ mas, and in the Cnt\_Ha data at $62.8^{+2.1}_{-1.9}$ mas. The corresponding PAs are $(97.8\pm0.9)^\circ$, $(99.4^{+1.1}_{-1.5})^\circ$ and $(99.0^{+1.5}_{-1.6})^\circ$, respectively. Errors in the PA measurements also take into account the above mentioned uncertainty in the astrometric calibration of the instrument, which was added in quadrature to the PA error bars. 

As within the error bars all filters gave the same results, we combined them and found that HD142527 B is located at a projected separation of $62.8^{+2.1}_{-2.7}$ mas from the primary star ($9.9^{+0.3}_{-0.4}$ AU at $157.3\pm1.2$ pc) and has a PA of $(98.7\pm1.8)^\circ$. The final values result from calculating the arithmetic mean of all the values obtained from the three different datasets, while their errors are calculated identically to those for each single dataset.  

\begin{figure}[b!]
\centering
\includegraphics[width=\hsize]{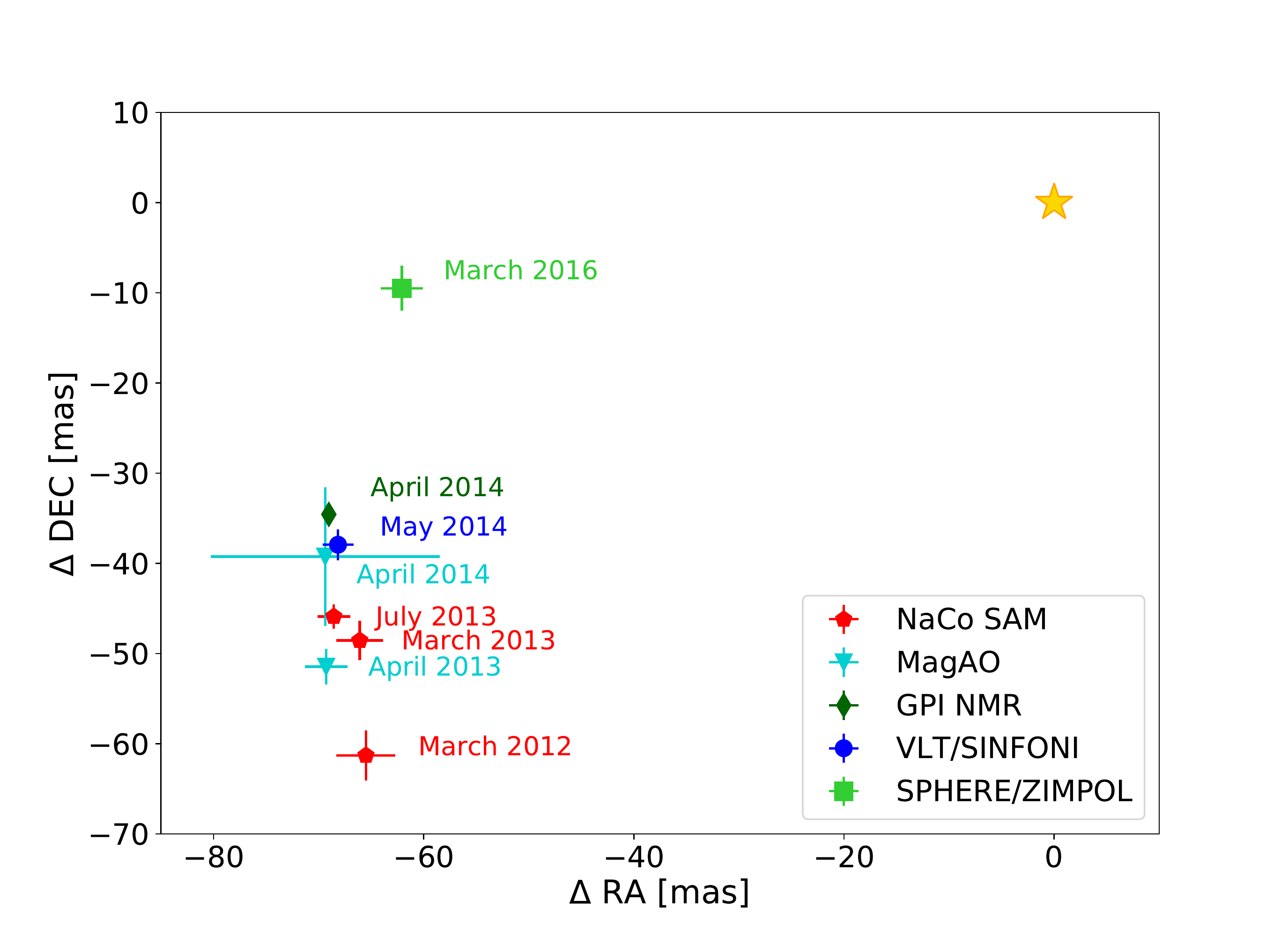}
\caption{Position of HD142527 B based on NaCo sparse aperture masking (red pentagons), MagAO (cyan triangles), GPI non-redundant masking (dark green diamonds) and VLT/SINFONI (blue circle) data from \cite{rodigas2014}, \cite {close2014}, \cite{lacour2016}, and \cite{Christiaens2018}, together with the SPHERE/ZIMPOL observation presented in this work (light green square). The position of HD142527 A is shown with the yellow star at coordinates (0,0).}
\label{fig:orbit}
\end{figure}

In Figure \ref{fig:orbit} we compare the positions previously estimated \citep{close2014, rodigas2014, lacour2016,Christiaens2018} and that resulting from our analysis. \cite{lacour2016} used a Markov chain Monte Carlo analysis to infer the orbital parameters of HD142527 B. Because the past detections were distributed over a relatively small orbital arc ($\sim 15^\circ$), it was difficult to constrain the parameters precisely. The high precision measurement added by our SPHERE/ZIMPOL data extends the arc to a range of $\sim30^\circ$. 
An updated orbital analysis is provided in \citep{Claudi2018}. Figure \ref{fig:orbit} shows that HD142527 B is clearly approaching the primary in the plane of the sky.

\begin{table*}[h!]
\caption{\label{tab:limits} Summary of our detection limits for each target. While for HD100546, HD169142, and MWC\,758 we consider the specific locations (separation and PA) of previously claimed companion candidates, we focused our analyses for HD135344B and TW Hya on separations related to disk gaps (hence no specific PA). Columns 5 and 6 give the mass and radius assumed for the accretion rate calculations, column 7 gives the contrast magnitude at the specific location and columns 8--11 report the values for the H$\alpha$ line flux, H$\alpha$ line luminosity, accretion luminosity, and mass accretion rate ignoring any possible dust around the companion.}
\centering
\small
\begin{tabular}{lllllllllll}
\hline\hline\noalign{\smallskip}
Target                  & Sep. & PA & Ref. & Mass & Radius & $\Delta$H$_\alpha$  &  $F^p_{H\alpha}$ & $L_{H\alpha}$ [$L_\odot$]         &  $L_\mathrm{acc}$ [$L_\odot$]  & $\dot{M}$ [$M_\odot\text{ yr}^{-1}$] \\ 
& [mas] & [$^\circ$] &  & [$M_J$] & [$R_J$] & [mag] & [erg/s/cm$^2$] & \\\hline
 \noalign{\smallskip}                    
                  HD135344B        & $180$ &  & (a) & 10.2\tablefootmark{(h)} & 1.6\tablefootmark{(j)} & $>9.8$  & $<3.8\times10^{-15}$  & $<2.0\times10^{-6}$        & $<3.7\times10^{-6}$ & $<2.4\times10^{-12}$  \\ \hline
 \noalign{\smallskip}                    
                  TW Hya        & $390$ &   & (b) & 2\tablefootmark{(k)} & 1.3\tablefootmark{(j)} & $>9.3$  & $<1.9\times10^{-14}$ & $<2.2\times10^{-6}$  & $<3.5\times10^{-6}$  & $<1.0\times10^{-11}$    \\ \hline
           \noalign{\smallskip} 
          \multirow{2}{*}{HD100546}             & $480\pm4$ & $8.9\pm0.9$ & (c) & 15\tablefootmark{(c)} & 2\tablefootmark{(j)} & $>11.4$  & $<1.1\times10^{-14}$  & $<4.7\times10^{-6}$  & $<1.1\times10^{-5}$ & $<6.4\times10^{-12}$  \\ \cline{2-11}
  \noalign{\smallskip}                  & $\sim140$ & $\sim133$ & (d) & 15\tablefootmark{(l)} & 2\tablefootmark{(j)} & $>9.3$  & $<7.9\times10^{-14}$  & $<3.3\times10^{-5}$  & $<2.0\times10^{-4}$ & $<1.1\times10^{-10}$  \\ \hline
  \noalign{\smallskip}
\multirow{2}{*}{HD169142}               & $\sim340$ & $\sim175$ & (e) & 0.6\tablefootmark{(e)} & 1.4\tablefootmark{(j)} & $>10.7$  & $<5.7\times10^{-15}$  & $<2.5\times10^{-6}$       & $<4.3\times10^{-6}$ & $<4.4\times10^{-11}$  \\ \cline{2-11}
  \noalign{\smallskip}                  & $156\pm32$ & $7.4\pm11.3$ & (f) &10\tablefootmark{(f)} & 1.7\tablefootmark{(j)} & $>9.9$  & $<1.2\times10^{-14}$  & $<5.2\times10^{-6}$  & $<1.3\times10^{-5}$ & $<7.6\times10^{-11}$  \\ \hline
  \noalign{\smallskip}
MWC 758       & $111\pm4$ & $162\pm5$ & (g) & 5.5\tablefootmark{(m)} & 1.7\tablefootmark{(n)} &$>9.4$  & $<1.4\times10^{-14}$  & $<1.2\times10^{-5}$  & $<4.3\times10^{-5}$ & $<5.5\times10^{-11}$  \\

\noalign{\smallskip}\hline\hline
\end{tabular}
\tablebib{\tablefoottext{a}{ \cite{Andrews2011}}; \tablefoottext{b}{\cite{garufi2013}}; \tablefoottext{c}{\cite{quanz2015}}; \tablefoottext{d}{\cite{Brittain2014}}; \tablefoottext{e}{\cite{osorio2014}}; \tablefoottext{f}{\cite{reggiani2014}}; \tablefoottext{g}{\cite{Reggiani2017}}; \tablefoottext{h}{\cite{maire2017}}, \tablefoottext{j}{AMES-Cond \citep{Allard2001, baraffe2003}}, \tablefoottext{k}{\cite{ruane2017}}, \tablefoottext{l}{\cite{Mendigutia2017}}, \tablefoottext{m}{\cite{Pinilla2015}}, \tablefoottext{n}{BT-Settl \citep{Allard2012}}.}
\end{table*}
\subsubsection{Photometry}
\label{sec:photometry_HD142527}
The Hessian matrix approach yields the contrasts between HD142527 A and B in every filter: $\Delta$N\_Ha $ = 6.3^{+0.2}_{-0.3} \text{ mag}$ in the narrow band, $\Delta$B\_Ha $ = 6.7\pm0.2 \text{mag}$ in the broad band, and $\Delta$Cnt\_Ha$ = 7.3 ^{+0.3}_{-0.2} \text{ mag}$ in the continuum filter. To quantify the brightness of the companion and not only its contrast with respect to the central star, we determined the flux of the primary in the multiple filters. We measured the count rate ($cts$) in the central circular region with radius $\sim1\farcs5$ in all frames of each stack and computed the mean and its uncertainty $\sigma/\sqrt{n}$, where $\sigma$ is the standard deviation of the count rate within the dataset and $n$ is the number of frames. No aperture correction was required because the same aperture size was used by \cite{schmid2017} to determine the zero points for the flux density for the three filters from photometric standard star calibrations. To estimate the continuum flux density we used their Equation 4 
\begin{equation}
F^*_\lambda(Cnt\_Ha)=cts\cdot10^{0.4\,(am\cdot k_1+m_{mode})}\cdot c_{zp}^{cont}(Cnt\_Ha),
\end{equation}
where $c_{zp}^{cont}(Cnt\_Ha)$ is the zero point of the Cnt\_Ha filter, $\textit{cts}=1.105\,(\pm0.001)\times10^5$ ct/s is the count rate measured from our data, $am=1.06$ is the average airmass, $k_1$ is the atmospheric extinction at Paranal \citep[$k_1(\lambda)=0.085$ mag/airmass for Cnt\_Ha, $k_1(\lambda)=0.082$ mag/airmass for B\_Ha and N\_Ha; cf.][]{Patat2011}, and $m_{mode}=-0.23$ mag is the mode dependent transmission offset, which takes into account the enhanced throughput of the R-band dichroic with respect to the standard gray beam splitter. The flux density of the primary star in the continuum filter $F^*_\lambda(Cnt\_Ha)$ was then used to estimate the fraction of counts in the line filters due to continuum emission via 
\begin{equation}
cts(F\_Ha)=\frac{F^*_\lambda(Cnt\_Ha)}{c_{zp}^{cont}(F\_Ha)}\times10^{-0.4(am\cdot k_1+m_{mode})},
\end{equation}
where $c_{zp}^{cont}(F\_Ha)$ is the continuum zero point of the H$\alpha$ filter used in the observations \citep[cf.][]{schmid2017}. During this step, we assumed that the continuum flux density was the same in the three filters. The continuum count rate was subtracted from the total count rate in B\_Ha and N\_Ha, $\text{cts}(B\_Ha)=1.631\,(\pm0.001)\times10^5$ ct/s and $\text{cts}(N\_Ha)=3.903\,(\pm0.003)\times10^4$ ct/s, leaving only the flux due to pure H$\alpha$ emission. These were used, together with Equation (1) with line zero points, to determine the pure H$\alpha$ line fluxes (see fifth column in Table \ref{tab:Ha_star}). For each filter, the continuum flux density was multiplied by the filter equivalent width, and the flux contribution from line emission was added for the line filters. As in \cite{sallum2015}, we assumed the B object to have the same extinction as A, ignoring additional absorption from the disk. Indeed, we considered an extinction of $A_V=0.05$ mag \citep{Fairlamb2015} and, interpolating the standard reddening law of \cite{Mathis1990} for $R_V=3.1$, we estimated the extinction at $\sim650$ nm to be A$_{H\alpha}=0.04$ mag. The stellar flux was found to be $6.1\,\pm0.2\times 10^{-11}$ erg/s/cm$^2$ in the Cnt\_Ha filter, $9.7\,\pm0.8\times10^{-11}$ erg/s/cm$^2$ in the B\_Ha filter and $3.0\,\pm0.8\times10^{-11}$ erg/s/cm$^2$ in the N\_Ha filter (see Table \ref{tab:Ha_star}).\\
With the empirically estimated contrasts, we calculated the companion flux, i.e., line plus continuum emission or continuum only emission, in the three filters as follows:  
$$F^p_{Cnt\_Ha}=7.4^{+1.4}_{-2.1}\times10^{-14}\text{ erg/s/cm}^2,$$ $$F^p_{B\_Ha}=2.0\,\pm0.4\times10^{-13}\text{ erg/s/cm}^2,$$ $$F^p_{N\_Ha}=9.1^{+3.5}_{-2.9}\times10^{-14}\text{ erg/s/cm}^2.$$\\
We note that the contrast we calculated in the continuum filter is very similar to that obtained by \cite{close2014} of $\Delta \text{mag} = 7.5\pm0.25$ mag. The direct estimation of the brightness of the primary in each individual ZIMPOL filter led to a larger difference when comparing the companion's apparent magnitude in our work ($m^B_{Cnt\_Ha}=15.4\pm0.2$ mag) with that from \cite{close2014} ($m^B_{\text{Close}}=15.8\pm0.3$ mag). Such values are possibly consistent within the typical variability of accretion of the primary and secondary at these ages. However, given the different photometry sources and filters used for the estimation of the stellar flux densities in the two works, the results cannot be easily compared.

\subsubsection{Accretion rate estimates}
\label{sec:accretion_HD142527}
The difference between the flux in the line filters and the continuum filter (normalized to the H$\alpha$ filter widths) represents the pure H$\alpha$ line emission for which we find for HD142527 B $f^{line}_{B\_Ha}=1.0^{+0.5}_{-0.4}\times10^{-13}$ erg/s/cm$^2$ and  $f^{line}_{N\_Ha}=7.6^{+3.5}_{-2.9}\times10^{-14}$ erg/s/cm$^2$, respectively. The line flux is then converted into a line luminosity multiplying it by the GAIA distance squared (see Table \ref{tab_stars}), yielding $L_{B\_Ha}=7.7^{+4.0}_{-3.6}\times10^{-5}\,L_\odot$ and $L_{N\_Ha}=6.0^{+2.8}_{-2.4}\times10^{-5}\,L_\odot$. We then estimated the accretion luminosity with the classical T Tauri stars (CTTS) relationship from \cite{rigliaco2012}, in which the logarithmic accretion luminosity grows linearly with the logarithmic H$\alpha$ luminosity
\begin{equation}
\log(L_\mathrm{acc}) = b+a\log(L_{H\alpha}), 
\end{equation}
and $a = 1.49\pm0.05$ and $b=2.99 \pm 0.16$ are empirically determined. We calculated the accretion luminosity for both datasets, yielding $L^\mathrm{acc}_{B\_Ha}=7.3^{+6.8}_{-6.4}\times10^{-4} L_\odot$ and $L^\mathrm{acc}_{N\_Ha}=5.0^{+4.4}_{-4.0}\times10^{-4} L_\odot$.\\
Following \citet{gullbring1998} we finally used 
\begin{equation}
\dot{M}_\mathrm{acc}=\left(1-\frac{R_c}{R_{in}}\right)^{-1} \frac{L_\mathrm{acc}R_c}{GM_c}\sim1.25\,\frac{L_\mathrm{acc}R_c}{GM_c}
\label{eq:accretion_rate}
\end{equation}
to constrain the mass accretion rate. The quantity $G$ is the universal gravitational constant, and $R_c$ and $M_c$ are the radius and mass of the companion, respectively. Assuming that the truncation radius of the accretion disk $R_{in}$ is $\sim5R_c$, we obtain $\left(1-\frac{R_c}{R_{in}}\right)^{-1}\sim1.25 $. For the companion mass and radius, two different sets of values were considered: \cite{lacour2016}  fitted the SED of HD142527 B with evolutionary models \citep{baraffe2003} and calculated $M_c=0.13\pm0.03\;M_\odot$ and $R_c=0.9\pm0.15\;R_\odot$, while \cite{Christiaens2018} estimated from H+K band VLT/SINFONI spectra $M_c=0.34\pm0.06\;M_\odot$ and $R_c=1.37\pm0.05\;R_\odot$, in the presence of a hot circumstellar environment\footnote{They considered two different cases in which the companion may or may not be surrounded by a hot environment contributing in $H$+$K$. Because of the presence of accreting material shown in this work, we decided to consider the first case.}. The accretion rates obtained from the H$\alpha$ emission line are $\dot{M}_{B\_Ha}=2.0^{+2.0}_{-1.9}\times 10^{-10}M_\odot/\text{yr}$ and $\dot{M}_{N\_Ha}=1.4^{+1.3}_{-1.2}\times 10^{-10}M_\odot/\text{yr}$ in the first case and $\dot{M}_{B\_Ha}=1.2\pm1.1\times 10^{-10}M_\odot/\text{yr}$ and $\dot{M}_{N\_Ha}=0.8\pm0.7\times 10^{-10}M_\odot/\text{yr}$ in the second case. 
Some H$\alpha$ flux loss from the instrument when the N\_Ha filter is used might explain the lower value of $\dot{M}_{N\_Ha}$ compared to $\dot{M}_{B\_Ha}$. Indeed, according to Figure 2 and Table 5 from \cite{schmid2017}, the N\_Ha filter is not perfectly centered on the H$\alpha$ rest wavelength, implying that a fraction of the flux could be lost, in particular if the line profile is asymmetric. Moreover, high temperature and high velocities of infalling material cause H$\alpha$ emission profiles of CTTS to be broad \citep{Hartmann1994,White_Basri2003}. Also, line broadening from the rotation and line shift of the object due to possible radial motion might be important, even though it is not expected to justify the $\sim$40\% H$\alpha$ flux difference of HD142527B. We argue, therefore, that with the available data it is very difficult to estimate the amount of line flux lost by the N\_Ha filter, and that the value given by the B\_Ha filter is expected to be more reliable, since all line emission from the accreting companion is included.\\
As shown in PDI images from \cite{Avenhaus2017}, dust is present at the separation of the secondary possibly fully embedding the companion or in form of a circumsecondary disk. During our calculations, we neglected any local extinction effects due to disk material. It is therefore possible that on the one hand some of the intrinsic H$\alpha$ flux gets absorbed/scattered and the actual mass accretion rate is higher than that estimated in this work; on the other hand, the material may also scatter some H$\alpha$ (or continuum) emission from the central star, possibly contributing in very small amounts to the total detected flux.\\
Although the results obtained in this work are on the same order of magnitude as those obtained by \cite{close2014}, who derived a rate of $6\times10^{-10}\,M_\odot \text{ yr}^{-1}$, it is important to point out some differences in the applied methods. Specifically, \cite{close2014} used the flux estimated in the H$\alpha$ filter to calculate $L_{H\alpha}$, while we subtracted the continuum flux and considered only the H$\alpha$ line emission. Moreover, we combined the derived contrast with the stellar flux in the H$\alpha$ filters obtained from our data, while \cite{close2014} used the $R$-band magnitude of the star. As HD142527 A is also accreting and therefore emitting H$\alpha$ line emission, this leads to a systematic offset. 
Finally, \cite{close2014} used the relationship found by \cite{Fang2009} and not that from \cite{rigliaco2012}, leading to a difference in the $L_\mathrm{H\alpha}-L_\mathrm{acc}$ conversion.

\subsection{HD135344 B}
Visual inspection of the final PSF-subtracted ADI images of HD135344B showed a potential signal north to the star. Given the weakness of the signal and the low statistical significance, we analyze and discuss it further in Appendix \ref{app:HD135344B_companion}. \\
In Figure \ref{fig:HD135344B_contrast_curves} we plot the contrast curves obtained as explained in section \ref{sec:setup_performance} using the N\_Ha and the Cnt\_Ha datasets and applying ASDI. In addition to the $1\farcs08\times1\farcs08$ images we also examined $2\farcs88\times2\farcs88$ images to search for accreting companions beyond the contrast limited region and beyond the spiral arms detected on the surface layer of the HD135344 B circumstellar disk. However, no signal was detected. We paid special attention to the separations related to the reported disk cavities \citep{Andrews2011, garufi2013}. We chose to investigate specifically the cavity seen in scattered light at $0\farcs18$. The outer radius of the cavity seen in millimeter continuum is larger, but small dust grains are expected to be located inside of this radius increasing the opacity and making any companion detection more difficult. Neglecting the small inclination \citep[$i\sim11^\circ$,][]{Lyo2011}, the disk is assumed to be face-on and the contrast value given by the curve of Figure \ref{fig:HD135344B_contrast_curves} at $0\farcs18$ is considered ($\Delta$N\_Ha = 9.8 mag). We derived the H$\alpha$ flux from the star in the N\_Ha filter as presented in section \ref{sec:photometry_HD142527} using the stellar flux values for the different filters given in Table \ref{tab:Ha_star}, and calculated the upper limits for the companion flux, accretion luminosity, and mass accretion rate following Section \ref{sec:photometry_HD142527} and Section \ref{sec:accretion_HD142527}. The accretion rate is given by Equation \ref{eq:accretion_rate}, assuming a planet mass of $M_c=10.2\,M_J$, the maximum mass that is nondetectable at those separations according to the analysis of \cite{maire2017}. Being consistent with their approach, we then used AMES-Cond\footnote{AMES-Cond and BT-Settl models used through the paper where downloaded on Feb. 06, 2018, from https://phoenix.ens-lyon.fr/Grids/AMES-Cond/ISOCHRONES/ and https://phoenix.ens-lyon.fr/Grids/BT-Settl/CIFIST2011\_2015/ISOCHRONES/, respectively.} evolutionary models \citep{Allard2001, baraffe2003} to estimate the radius of the object $R_c=1.6\,R_J$ based on the age of the system. All values, sources, and models used are summarized in Table \ref{tab:Ha_star} and in Table \ref{tab:limits} together with all the information for the other objects. The final accretion rate upper limit has been calculated to be $<2.4\times10^{-12}\,M_\odot\text{ yr}^{-1}$ at an angular separation of $0\farcs18$, i.e., the outer radius of the cavity seen in scattered light.
\begin{figure}[b!]
\centering
\includegraphics[width=\hsize]{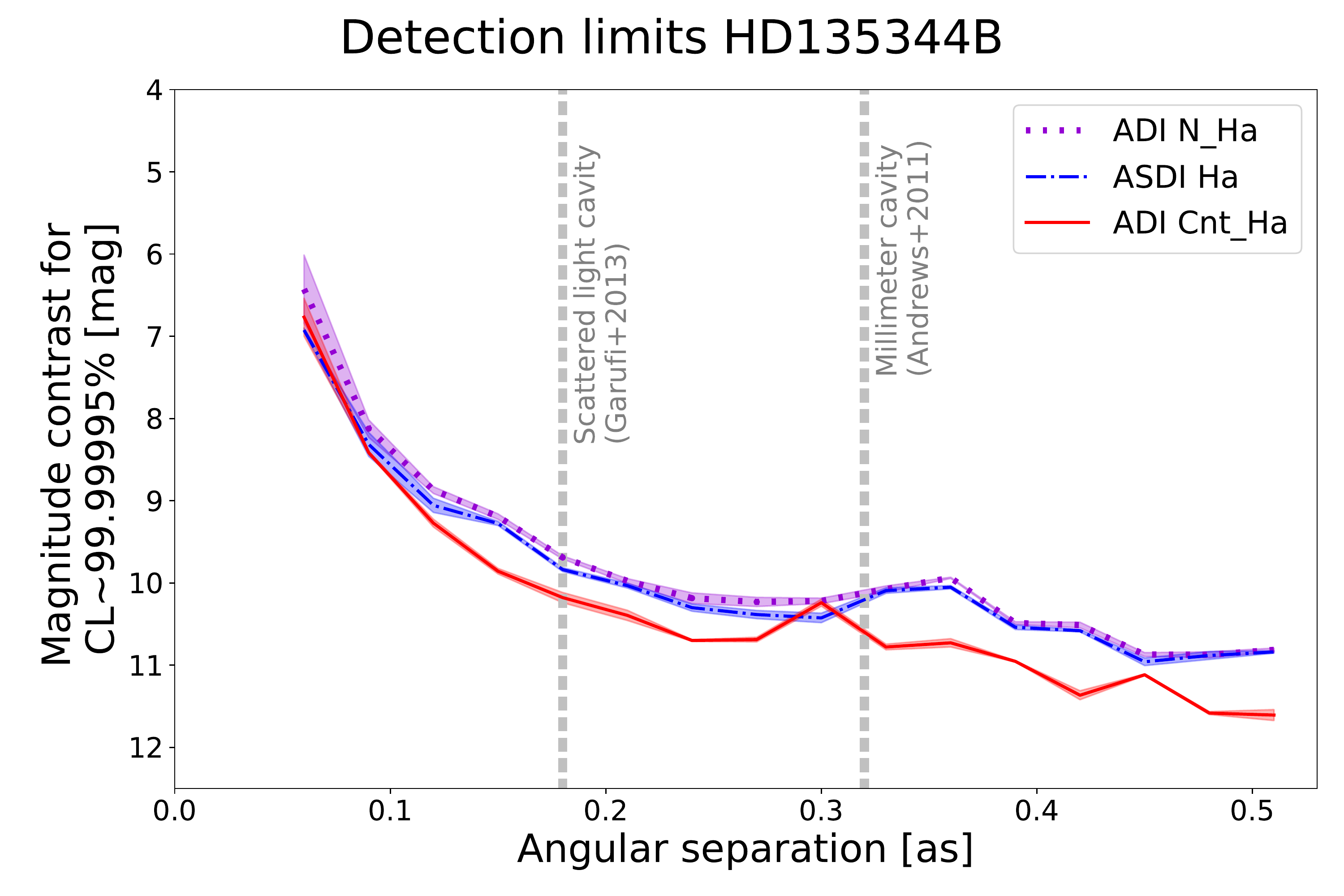}
\caption{Contrast curves for HD135344 B. The vertical lines indicate the outer radii of the cavities in small and large dust grains presented in \cite{garufi2013} and \cite{Andrews2011}, respectively.}
\label{fig:HD135344B_contrast_curves}
\end{figure}

\begin{figure*}[t!]
\centering
\includegraphics[width=\hsize]{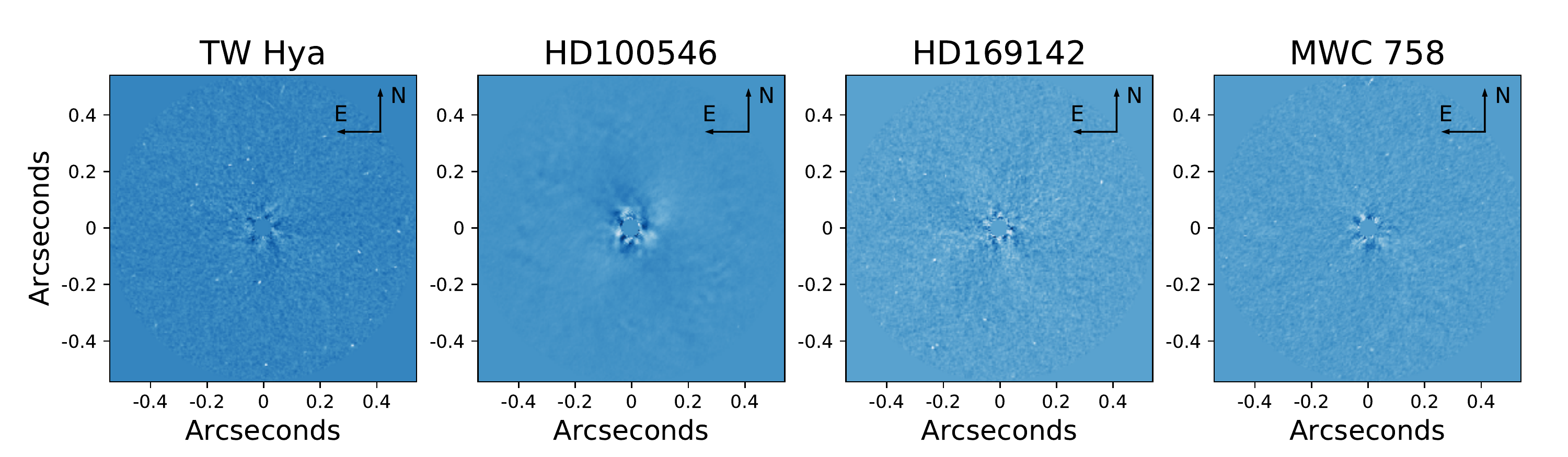}
\caption{Final PSF subtracted ADI images of TW Hya, HD100546, HD169142, and MWC\,758. We applied a central mask with radius 32.4 mas and 18 PCs were removed. No companion candidates were detected. All images have a linear, but slightly different, color scale.}
\label{fig:others}
\end{figure*}
\subsection{TW Hya}
\begin{figure}[t!]
\centering
\includegraphics[width=\hsize]{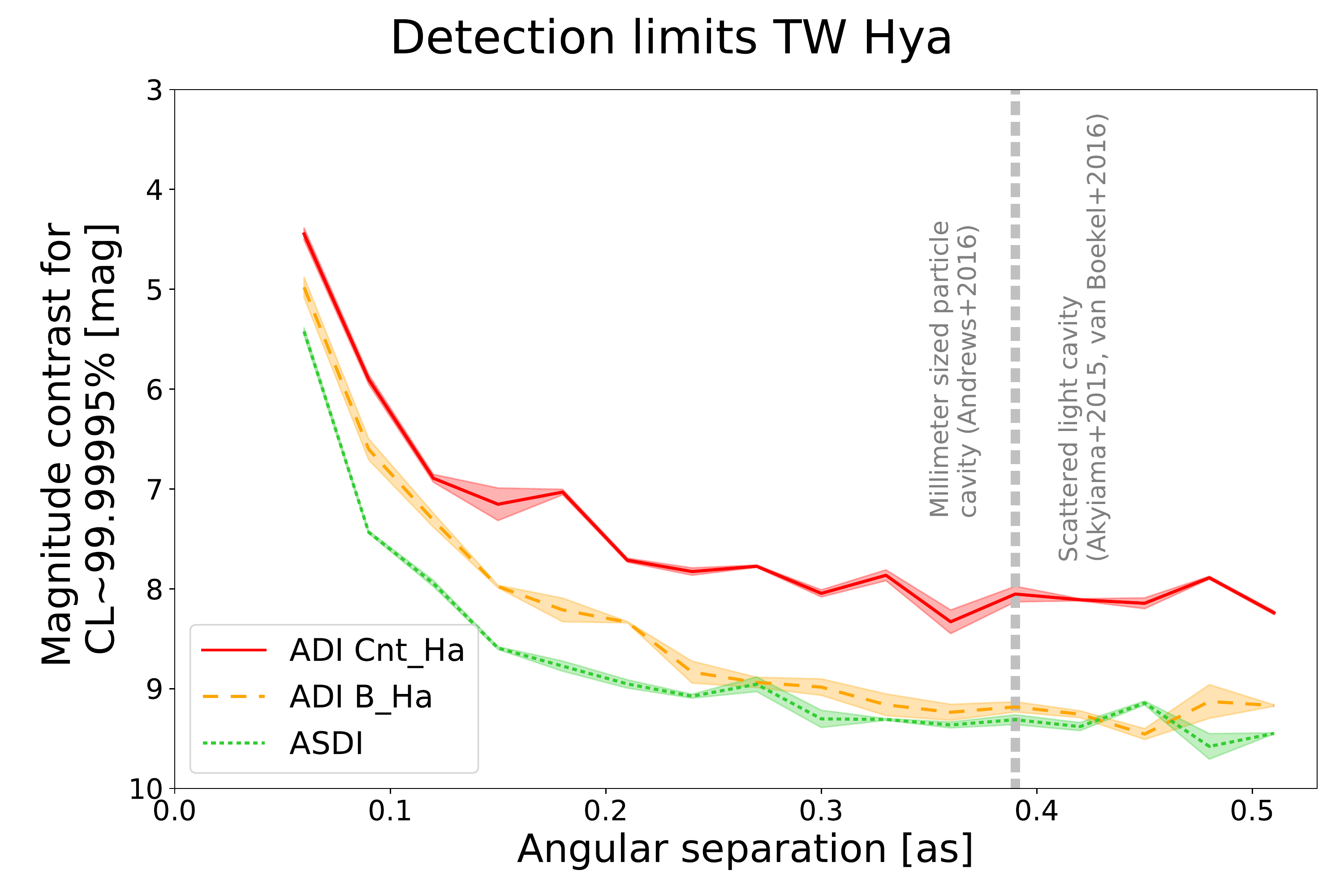}
\caption{Contrast curves for TW Hya. The vertical line indicates the gap at $0\farcs39$ detected in both scattered light \citep{Akiyama2015,vanboekel2017} and submillimeter continuum \citep{Andrews2016}.}
\label{fig:TWHya_contrast_curves}
\end{figure}

The TW Hya dataset does not show any point source either in the $1\farcs08\times1\farcs08$ images (see Figure \ref{fig:others}) or in the $2\farcs88\times2\farcs88$ images, which are large enough to probe all the previously reported disk gaps. The final contrast curves are shown in Figure~\ref{fig:TWHya_contrast_curves}. We also looked specifically at detection limits within the gaps observed by \cite{vanboekel2017} and focused in particular on the dark annulus at 20 AU ($0\farcs39$) from the central star, which has a counterpart approximately at the same position in 870 $\mu$m dust continuum observations \citep{Andrews2016}

Since the circumstellar disk has a very small inclination, we considered the disk to be face-on and assumed the gaps to be circular. At $0\farcs39$, planets with contrast lower than 9.3 mag with respect to TW Hya would have been detected with the ASDI technique (cf. Figure~\ref{fig:TWHya_contrast_curves}). This value was then combined with the stellar flux calculated as described in section \ref{sec:photometry_HD142527}, to obtain the upper limit of the companion flux in the B\_Ha filter. This yielded $\dot{M}<1.0\times10^{-11}\,M_\odot\text{ yr}^{-1}$ (see Table~\ref{tab:limits}) as the upper limit for the mass accretion rate based on our SPHERE/ZIMPOL dataset.

\subsection{HD100546}
\label{sec:Analysis_HD100546}
\begin{figure}[t!]
\centering
\includegraphics[width=\hsize]{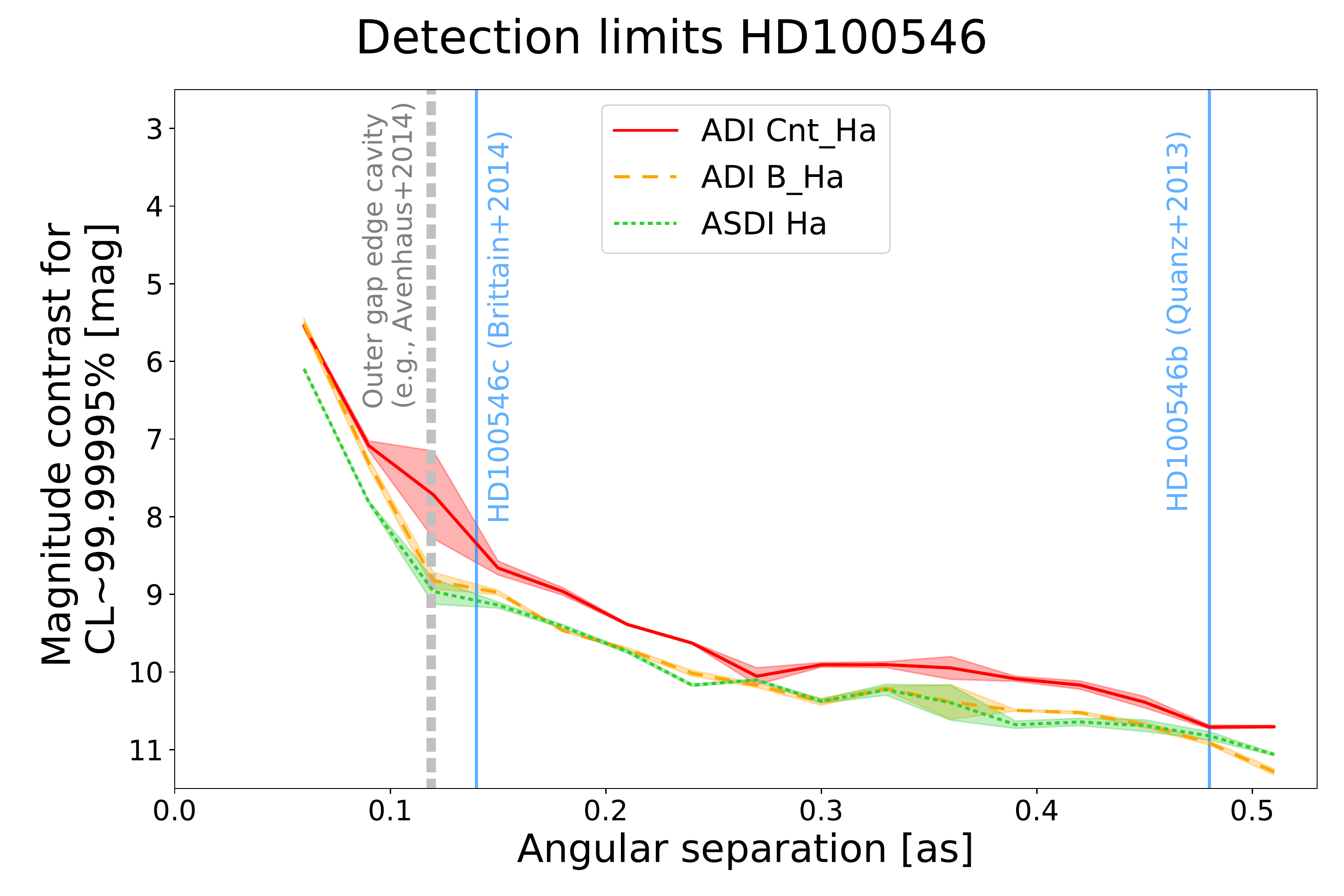}
\caption{Contrast curves for HD100546. The gray dashed vertical line shows the separation of the outer gap edge cavity presented in \cite{Avenhaus2014}, while the solid blue lines indicate the separations of the forming planet candidates around HD100546 \citep{Quanz2013_discovery,Brittain2014}.}
\label{fig:HD100546_contrast_curves}
\end{figure}

The HD100546 dataset suffered from rather unstable and varying observing conditions, which resulted in a large dispersion in the recorded flux (see Figure \ref{fig:HD100546_frame_sel} in Appendix~\ref{App_3}). 
We hence selected only the last 33\% of the observing sequence, which had relatively stable conditions, for our analysis (see Appendix~\ref{App_3}). 
The H$\alpha$ data did not confirm either of the two protoplanet candidates around HD100546 (see Figure \ref{fig:others}) and we show the resulting detection limits in Figure \ref{fig:HD100546_contrast_curves}. 

In order to investigate the detection limits at the positions of the protoplanet candidates, we injected artificial planets with increasing contrast starting from $\Delta$B\_Ha = 8.0 mag until the signal was no longer detected with a CL of at least 99.99995\%, and we repeated the process subtracting different numbers of PCs (from 10 to 30). At the position where \cite{quanz2015} claimed the presence of a protoplanetary companion, we would have been able to detect objects with a contrast lower than 11.4 mag (using PC=14 and the ADI reduction). Consequently, if existing, a 15 M$_J$ companion \citep{quanz2015} located at the position of HD100546 b must be accreting at a rate $<6.4\times10^{-12}\,M_\odot\text{ yr}^{-1}$ in the framework of our analysis and assuming no dust is surrounding the object. We note that, in comparison to the accretion luminosity $L_\mathrm{acc}$ estimated by \cite{Rameau2017}, our upper limit is one order of magnitude lower (cf. Table \ref{tab:limits}).
For the position of HD100546\,c, we used the orbit given in \cite{Brittain2014} to infer the separation and PA of the candidate companion at the epoch of our observations, i.e., $\rho\simeq0\farcs14$ and $\text{PA}\simeq 133^\circ$.  At this position our data reach a contrast of 9.3 mag (using PC=14 on the continuum-subtracted dataset), implying an upper limit for the companion flux in the H$\alpha$ filter of $7.9\times10^{-14}$ erg/s/cm$^2$ and a mass accretion rate $<1.1\times10^{-10}\,M_\odot\text{ yr}^{-1}$. This puts $\sim2$ orders of magnitude stronger constraints on the accretion rate of HD100546\,c than the limits obtained from the polarimetric H$\alpha$ images presented in \cite{Mendigutia2017} for a $15\,M_J$ planet. We note that owing to its orbit, HD100546\,c is expected to have just disappeared or to disappear quickly behind the inner edge of the disk \citep{Brittain2014}. Therefore, extinction could play a major role in future attempts to detect this source. 

\subsection{HD169142}
\begin{figure}[t!]
\centering
\includegraphics[width=\hsize]{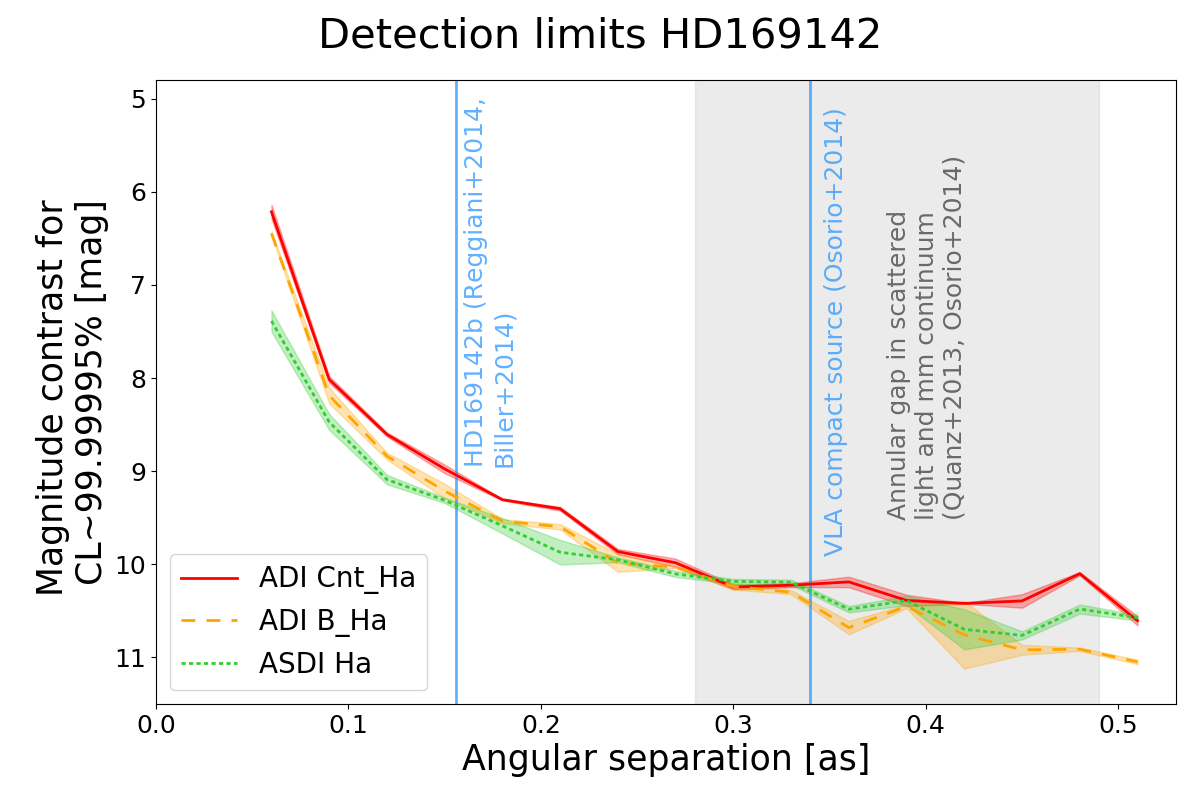}
\caption{Contrast curves for HD169142. The shaded region represents the annular gap observed in scattered light \citep{quanz2013} and in millimeter continuum \citep{osorio2014}. The blue vertical lines represent the separation of the companion candidates \citep{reggiani2014,biller2014,osorio2014}.}
\label{fig:HD169142_contrast_curves}
\end{figure}

We analyzed the data with ADI and ASDI reductions (see Figure \ref{fig:others} for the ADI image). The latter  was particularly interesting in this case because the stellar flux density in the continuum and H$\alpha$ filter is very similar and the continuum subtraction almost annihilated the flux from the central PSF, indicating that the central star has limited to no H$\alpha$ line emission (cf. Table \ref{tab:Ha_star} and see \cite{Grady2007}). We calculated the detection limits as explained in section \ref{sec:setup_performance} for both filters for a confidence level of 99.99995\%, as shown in Figure~\ref{fig:HD169142_contrast_curves}.\\

We investigated with particular interest the positions of the candidates mentioned in Section \ref{sec:sample} and derived specific detection limits at their locations, independent from the azimuthally averaged contrast curve. At the position of the compact source found by \cite{osorio2014} (we call this potential source HD169142\,c), our data are sensitive to objects 10.7 mag fainter than the central star (obtained by subtracting 16 PCs with ASDI reduction). At the position of HD169246\,b \citep{reggiani2014,biller2014} an object with a contrast as large as 9.9 mag could have been detected (PC=19; ASDI). For the compact source from \cite{osorio2014} we found $\dot{M}<4.4\times10^{-11}\,M_\odot\text{ yr}^{-1}$. 
Similarly, for the object detected by \cite{biller2014} and \cite{reggiani2014}\footnote{Within the uncertainties in the derived positions, these objects are indistinguishable and hence we assume it is the same candidate.} we found an upper limit for the mass accretion rate of $\dot{M}<7.6\times10^{-11}\,M_\odot\text{ yr}^{-1}$.

\subsection{MWC\,758}
Our analysis of the SPHERE/ZIMPOL images did not show an H$\alpha$ counterpart to the MWC\,758  companion candidate detected by \citet{Reggiani2017} as shown in Figure \ref{fig:others}. This is consistent with the recently published results from \cite{Huelamo2018}. Nonetheless, we provide a detailed analysis and discussion of the same MWC\,758 data to allow a comparison with the other datasets. 

In Figure \ref{fig:MWC758_contrast_curves} we show the detection limits obtained with ADI for the B\_Ha and Cnt\_Ha dataset, and the results of the ASDI approach. At separations larger than $0\farcs25$, companions with a contrast smaller than 10 mag could have been detected. At the specific position of the candidate companion\footnote{For our analysis we considered the position obtained from the first dataset in \cite{Reggiani2017} because the observing date was close to the epoch of the H$\alpha$ observations.} we can exclude objects with contrasts lower than 9.4 mag (obtained subtracting 15 PCs using ASDI).

To explain the presence of a gap in dust-continuum emission without a counterpart in scattered light, a steady replenishment of $\mu$m-sized particle is required, which implies that a companion in the inner disk should not exceed a mass of $M_c=5.5\,M_J$ \citep{Pinilla2015, Reggiani2017}. In line with the analysis of \cite{Reggiani2017}, we used the BT-Settl model to estimate the radius of the companion and we derived an upper limit for the mass accretion rate of $\dot{M}<5.5\times10^{-11}\,M_\odot\text{ yr}^{-1}$ (see Table~\ref{tab:limits}). Our analysis puts slightly stronger constraints on the mass accretion rate in comparison to that in \cite{Huelamo2018}.

\section{Discussion}
\label{sec:discussion}

\subsection{SPHERE/ZIMPOL as hunter for accreting planets}
The SPHERE/ZIMPOL H$\alpha$ filters allow for higher angular resolution compared to filters in the infrared regime and can, in principle, search for companions closer to the star. For comparison, a resolution element is 5.8 times smaller in the H$\alpha$ filter than in the $L'$ filter, meaning that the inner working angle (IWA) is smaller by the same amount so that closer-in objects could be observed, if bright enough\footnote{We note that SPHERE does not operate at similarly high Strehl ratios in the optical regime as it is able to do in the infrared.}. An instrument with similar capabilities is MagAO \citep{MagAO2014, MagAO2016}, but as the Magellan telescope has a primary mirror of 6.5 m diameter, it has a slightly larger IWA than SPHERE at the 8.2 m VLT/UT3 telescope. A direct comparison of the HD142527\,B detection shows that ZIMPOL reaches a factor $\sim2.5$ higher S/N in one-third of total integration time and field rotation of MagAO under similar seeing conditions, even if the companion is located $\gtrsim20$ mas closer to the star. 
The VAMPIRES instrument combined with Subaru/SCExAO will soon be a third facility able to perform H$\alpha$ imaging in SDI mode \citep{Norris2012}

\begin{figure}[t!]
\centering
\includegraphics[width=\hsize]{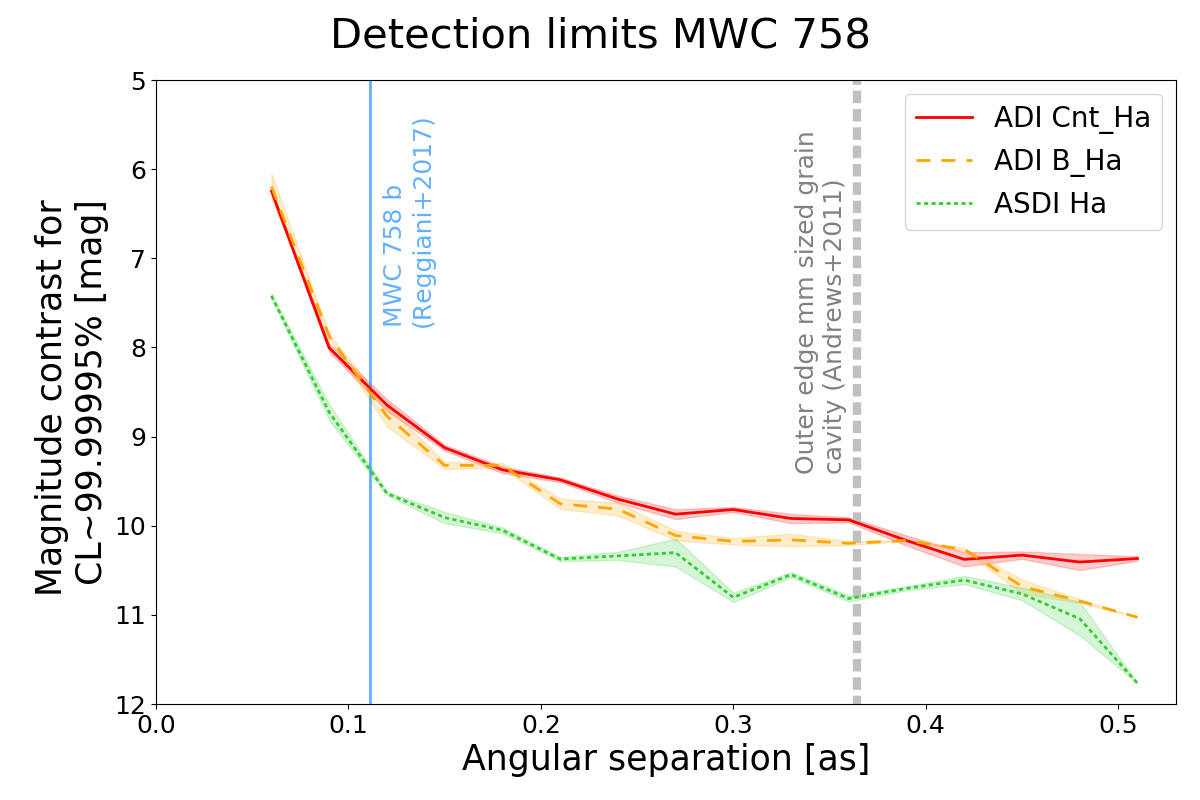}
\caption{Contrast curves for MWC\,758. The gray dashed line shows the outer edge of the dust cavity observed by \cite{Andrews2011}. The blue solid line indicates the separation at which \cite{Reggiani2017} found a candidate companion.}
\label{fig:MWC758_contrast_curves}
\end{figure}

\begin{figure}[t!]
\centering
\includegraphics[width=\hsize]{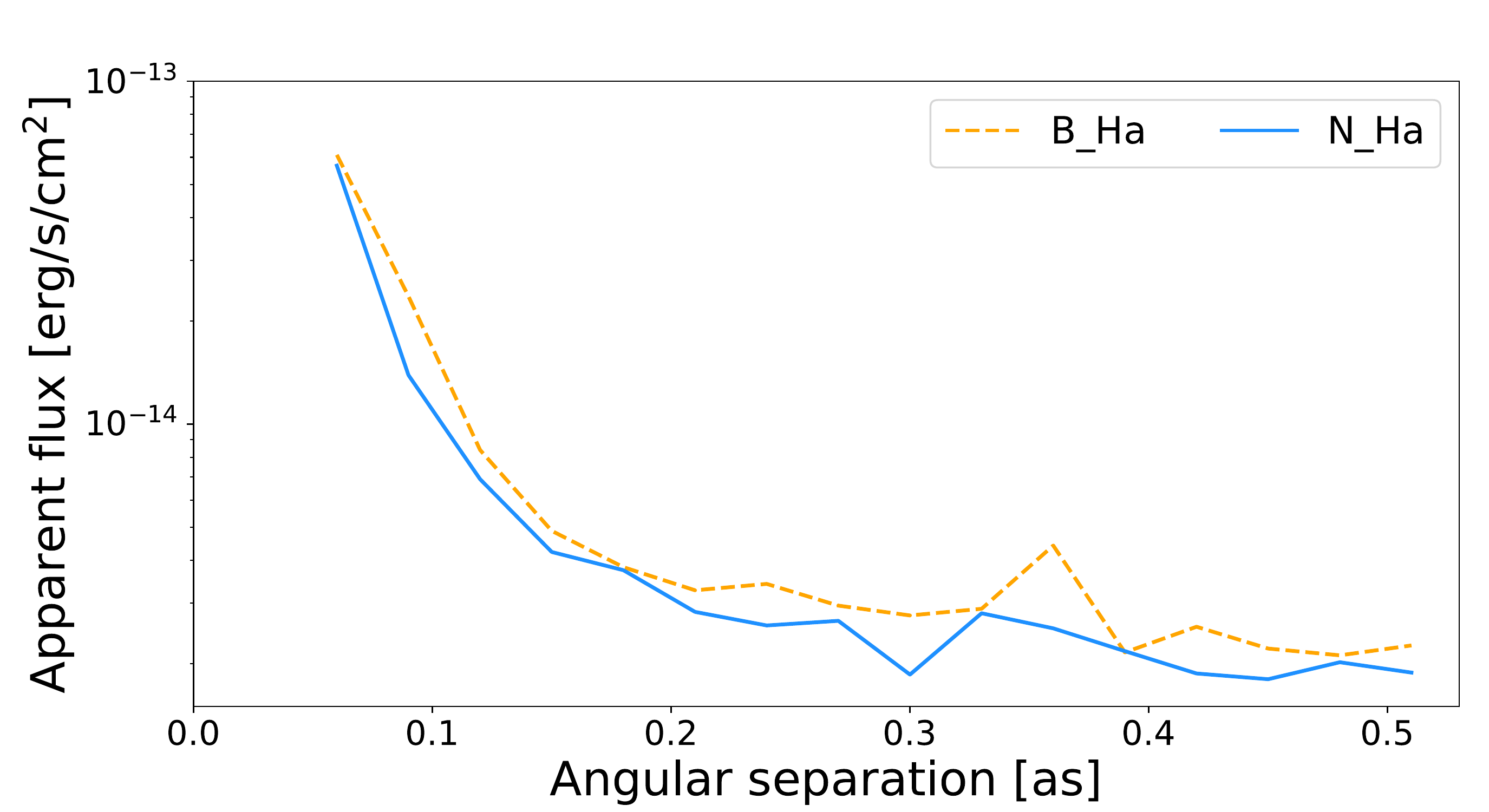}
\caption{Apparent flux detection limits as a function of the angular separation from HD142527 for both B\_Ha and N\_Ha filters.}
\label{fig:Broad_or_Narrow}
\end{figure}

In terms of detection performance using different filters and reduction techniques, we re-emphasize that the N\_Ha filter is more efficient in detecting H$\alpha$ signals in the contrast limited regime. The smaller filter width reduces the contribution of the continuum flux, which often dominates the signal in the B\_Ha filter, particularly for the central star. Hence, assuming the planetary companion emits only line radiation, the N\_Ha filter reduces the contamination by the stellar signal in the remaining speckles. Moreover, the subtraction of the stellar continuum from H$\alpha$ images reduces the speckles in both B\_Ha and N\_Ha filters. Hence, ASDI enhances the signal of potential faint companions, in particular at separations $<0\farcs3$ (cf. Figures \ref{fig:TWHya_contrast_curves}, \ref{fig:HD169142_contrast_curves}, \ref{fig:MWC758_contrast_curves}), where companions 0.7 mag fainter appear accessible in comparison to using simple ADI. ASDI should always be applied during the analysis of SPHERE/ZIMPOL H$\alpha$ data.
What remains to be quantified is how longer detector integration times (DITs) or the broad band filter could improve the detection limits in the background limited regime (i.e., $>0\farcs3$ where the contrast curves are typically flattening out) or for fainter natural guide stars. At these separations narrow band data can be detector read noise limited and the B\_Ha filter might be more suitable because of its higher throughput. However, as we show in Figure \ref{fig:Broad_or_Narrow}, it seems that at least for our HD142527 dataset this does not seem to be the case. Future studies conducted in both filters and on several objects are required to derive a more comprehensive understanding. Finding the sweetspot between longer integration times and the smearing of the PSF because of field rotation is also warranted. At least for the object considered in Figure \ref{fig:Broad_or_Narrow}, at large separations (usually $>0\farcs3$, in the background limited region) it is even possible to ignore completely ADI and simply apply field stabilized observations.

\subsection{Constraining planet accretion}
\begin{figure}[b!]
\centering
\includegraphics[width=\hsize]{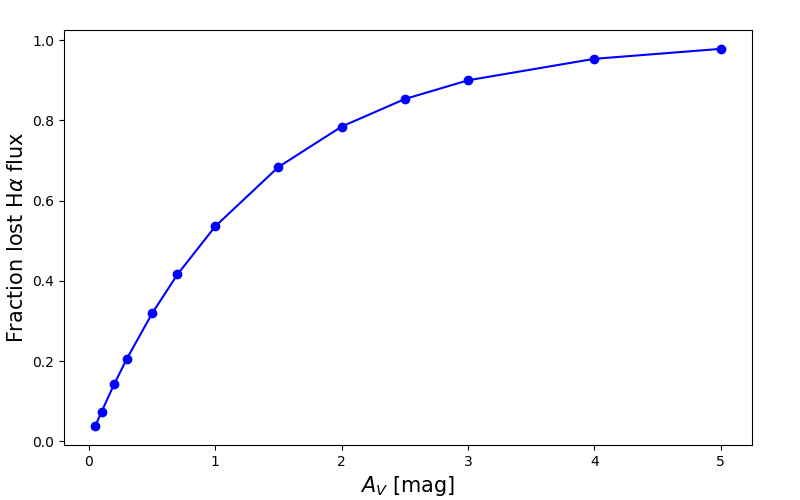}
\caption{Fraction of H$\alpha$ flux absorbed as a function of the disk extinction $A_V$ assuming the extinction law of \cite{Mathis1990} as explained in Section \ref{sec:photometry_HD142527}.}
\label{fig:extinction}
\end{figure}

For our mass accretion rate estimates of HD142527 B we assumed that 100\% of the H$\alpha$ flux originates from accretion processes involving circumstellar material. We note, however, that the values may be overestimated if we consider that chromospheric activity of the M star \citep{White_Basri2003, Fang2009} can also contribute to the measured line flux. Furthermore, as mentioned in Section \ref{sec:accretion_HD142527}, we warn that the narrow width of the N\_Ha filter might be too narrow to fully encompass all H$\alpha$ line emission from fast-moving, accreting material, and therefore the results may be underestimated. Finally, given the presence of dusty material at the projected position of HD142527
B \citep{Avenhaus2017}, H$\alpha$ flux might have been partially absorbed. It is beyond the scope of this paper to properly estimate a value for intrinsic extinction due to disk material and consider this value in the $\dot{M}$ estimation. Nevertheless, in Figure \ref{fig:extinction} we show the fraction of H$\alpha$ flux that is potentially lost because of extinction as a function of $A_V$, converted into $A_{H\alpha}$ as explained in Section \ref{sec:photometry_HD142527}. Only 2\% of the H$\alpha$ signal remains if the disk material causes an extinction of $A_V=5$ mag. This plot quantifies the impact of dust on the measured flux and the detectability of H$\alpha$ emission from embedded objects.

\begin{figure}[b!]
\centering
\includegraphics[width=\hsize]{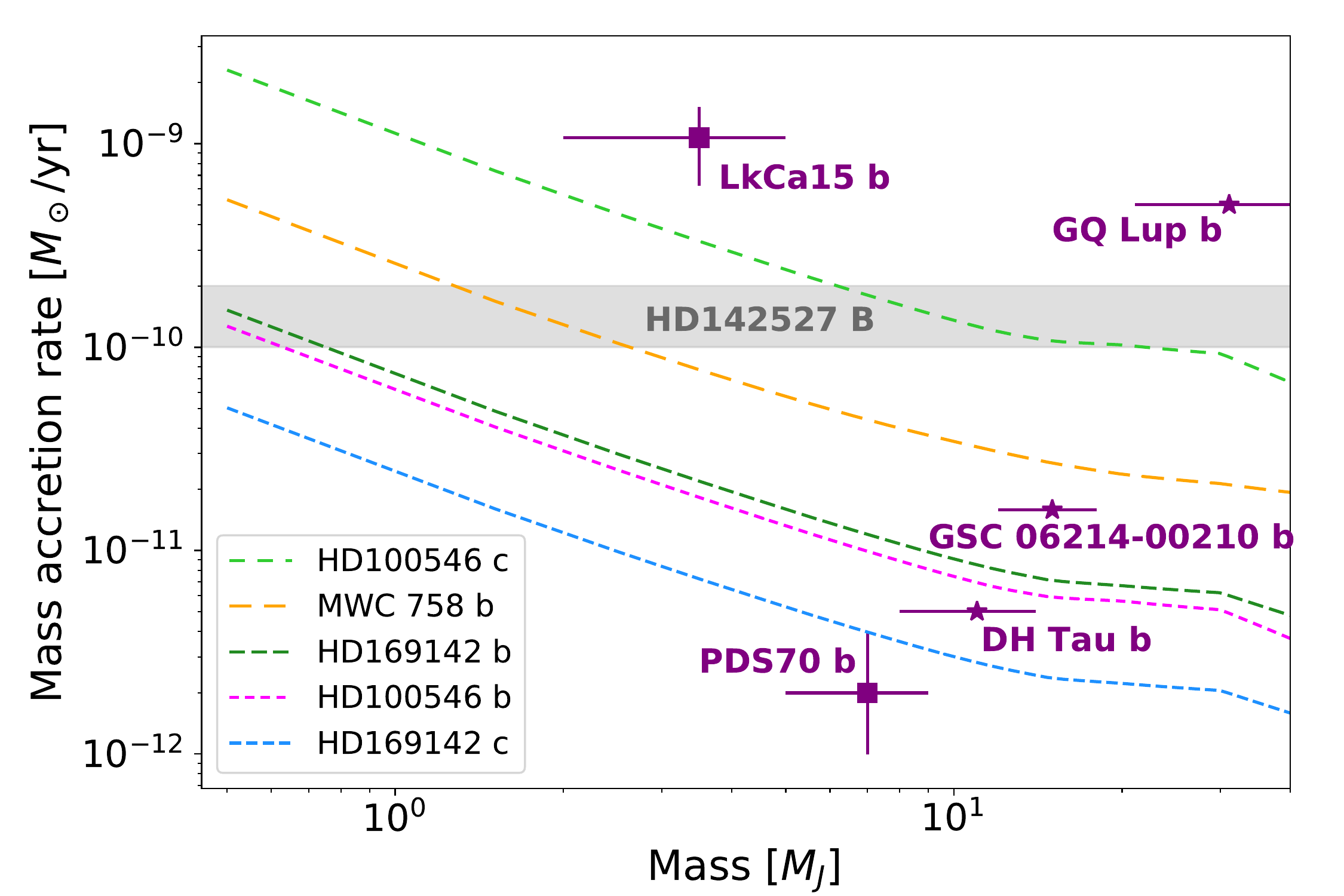}
\caption{Mass accretion rate upper limits as a function of the planetary mass for all the candidate forming planets investigated in this work. The violet stars represent the values reported in \cite{Zhou2014}, while the violet squares indicate PDS70\,b \citep{Wagner2018} and LkCa15\,b \citep{sallum2015}. The gray shaded area represents the mass accretion rate of HD142527 B and is shown for mass accretion rate comparison purposes only. Indeed, the mass of the object is much larger than what is reported on the x-axis of the plot.}
\label{fig:mass_accretion_rates}
\end{figure}

For the other five objects studied in this work we were not able to detect any clear accretion signature located in the disks. Therefore, our data were not able to support the scenario in which protoplanets are forming in those disks. We put upper limits on the accretion luminosity and mass accretion rate. Two notes have to be made: (1) the fundamental quantities \emph{directly derived from the data} are $F_\mathrm{H\alpha}$ and $L_\mathrm{H\alpha}$; they should be used for future comparisons with other  datasets or objects; (2) the presented upper limits on $\dot{M}$ are only valid for an object with the mass and radius given in Table \ref{tab:limits}, while the $L_\mathrm{acc}$ upper limits refer to objects of any mass. In particular, assuming lower mass objects implies larger $\dot{M}$, as shown in Figure~\ref{fig:mass_accretion_rates}: on the y-axis the mass accretion rate upper limits decrease as a function of the companion mass, for which the corresponding radius was calculated using the evolutionary models reported in Table \ref{tab:limits} and assuming the age listed in Table \ref{tab_stars}. The plot highlights that the assumed mass of the companion may change the final $\dot{M}_{acc}$ by more than one order of magnitude. Moreover, we overplot in violet the mass accretion rates of the three objects presented in \citet[][see also Section \ref{sec:discussion_objects}]{Zhou2014} as well as LkCa15\,b and PDS70\,b \citep{sallum2015,Wagner2018}, and in gray the range of mass accretion rates for HD142527.

We stress that, similar to HD142527 B, we always assumed that the flux limit is completely due to H$\alpha$ line emission without any contribution from continuum or chromospheric activity. Furthermore, for our analysis we always neglected intrinsic extinction effects from disk material, which likely weaken the signal. In particular, at locations where no gap in small dust grains has been identified the extinction $A_{H\alpha}$ can be significant (see Figure \ref{fig:extinction}). Models and precise measurements of the dust content in the individual disks would be required to properly include local extinction into our analysis. 
Finally, investigating the H$\alpha$ luminosity upper limits for the specific positions as a function of the separation from the central star, it can be noticed that the constraints are stronger at larger separations. The only exception is HD100546, for which higher upper limits were achieved. The combination of suboptimal weather conditions, under which the dataset was taken, and the small field rotation of the subsample analyzed in this work made those limits worse. A more stable dataset with larger field rotation should provide more constraining limits.
\subsection{Comparison with other objects}
\label{sec:discussion_objects}
The accretion rate of HD142527 B is in good agreement with the mass accretion rates found in \cite{rigliaco2012} for low-mass TTauri stars in the $\sigma$ Ori star-forming region ($5\times10^{-11}\,M_\odot\text{ yr}^{-1}<\dot{M}_{CTTS}<10^{-9}\,M_\odot\text{ yr}^{-1}$). A slightly broader mass accretion rate range was found by \cite{Alcala2014}, with $2\times10^{-12}\,M_\odot\text{ yr}^{-1}<\dot{M}_{CTTS}<4\times10^{-8}\,M_\odot\text{ yr}^{-1}$ in the Lupus star-forming region. 

\cite{Zhou2014} reported three very low-mass objects (GSC 06214-00210 b, GQ Lup b and DH Tau b), which exhibit H$\alpha$ emission from accretion. Those objects have separations of 100-350 AU from their parent stars and $\dot{M}\sim10^{-9}-10^{-11}\,M_\odot\text{ yr}^{-1}$ (see violet stars in Figure \ref{fig:mass_accretion_rates}). The accretion rates measured in the paper are of the same order as the limits we found in our work. At projected distances similar to those of the three objects mentioned above, ZIMPOL would have been able to observe and detect H$\alpha$ emitting companions. However, closer to the star in the contrast limited regime, our data would not have detected accretion processes occurring with $\dot{M}\lesssim10^{-11}\,M_\odot\text{ yr}^{-1}$.

\begin{figure}[t!]
\centering
\includegraphics[width=\hsize]{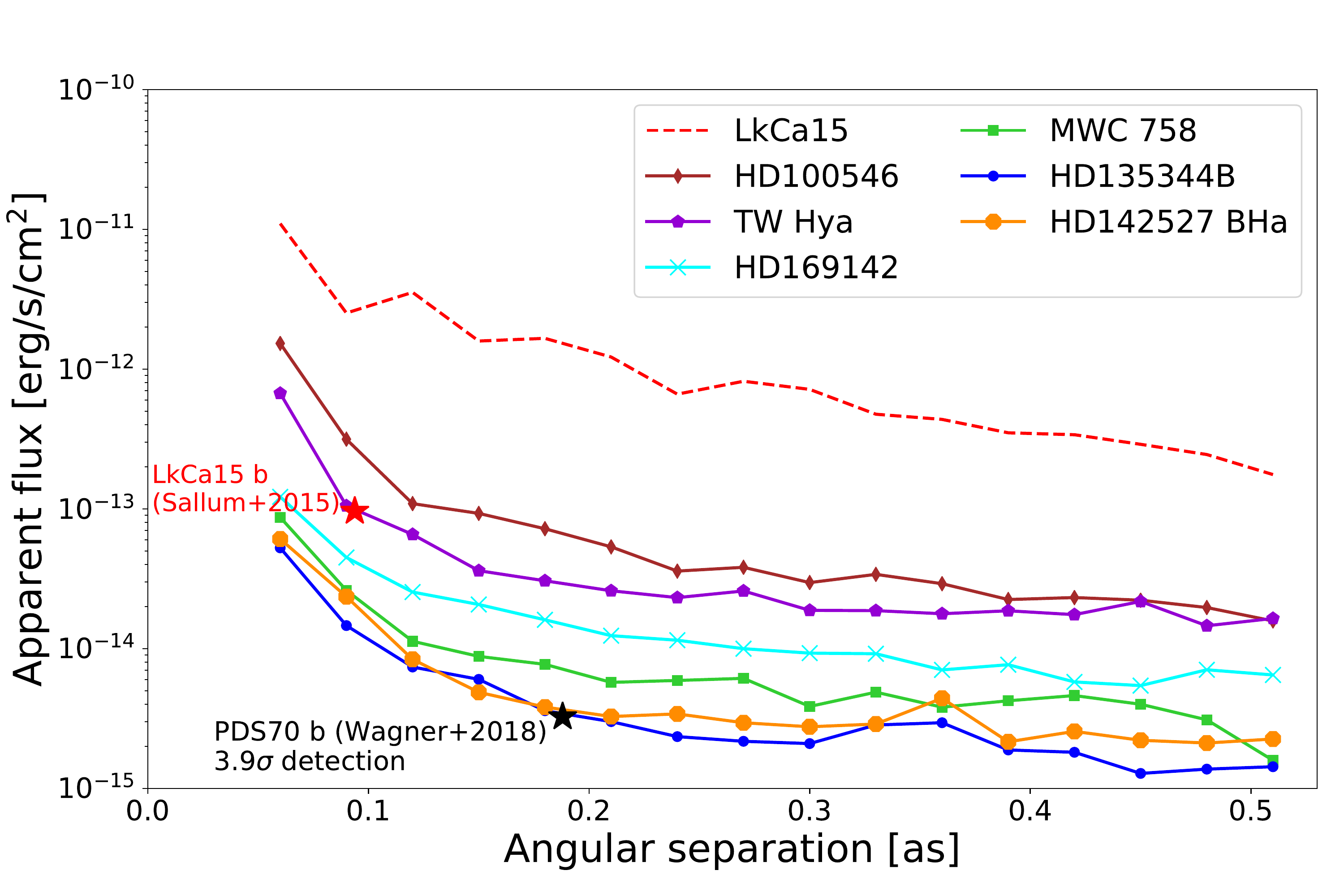}
\caption{Detection limits in apparent flux obtained for a 99.99995\% CL in this work, together with limits achieved with the available ZIMPOL dataset for LkCa15\,b (red dashed line) and the result presented in \cite{sallum2015} and \cite{Wagner2018}. A deeper dataset is required to redetect LkCa15\,b with ZIMPOL, but this detection is feasible.}
\label{fig:LkCa15_comparison}
\end{figure}

The mass accretion rate of PDS70\,b was estimated by \cite{Wagner2018} without considering any extinction effects and it is slightly lower than the limits we achieve for our sample (see violet square in Figure~\ref{fig:mass_accretion_rates} and black star in Figure \ref{fig:LkCa15_comparison}). The flux was calculated from the contrast in \cite{Wagner2018} assuming $R_{\text{PDS70\,b}}=11.7$ mag and estimating the MagAO H$\alpha$ filter widths assuming a flat SED\footnote{ \url{https://visao.as.arizona.edu/software_files/visao/html/group__reduction__users__guide.html\#visao_filters}}. In order to properly compare our limits and their H$\alpha$ detection, the same confidence levels should be considered. We therefore estimated the contrast limit for a CL corresponding to a $4\sigma$ detection for HD142527 at the separation of PDS70\,b, which was 0.3 mag lower than the limits corresponding to a CL of 99.99995\%. Hence, to bring all the contrast curves from Figure \ref{fig:LkCa15_comparison} to a 4$\sigma$ confidence level at $\sim0\farcs19$, a multiplication by a factor 0.76 is required. We note, however, that this scaling is just an approximation to provide a more direct comparison between the two studies. 

We also compared the H$\alpha$ line luminosity upper limits obtained from our ZIMPOL H$\alpha$ sample with that estimated by \cite{sallum2015} for LkCa15 b ($L_{H\alpha}\sim6\times10^{-5}\,L_\odot$). Our specific limits for the candidates around  HD169142, HD100546, and MWC 758 are slightly lower, but, except for HD100546 b and the compact source in HD169142 found by \cite{osorio2014}, of the same order of magnitude. LkCa15 itself was observed with SPHERE/ZIMPOL during the science verification phase in ESO period P96. We downloaded and analyzed the data, which were, however, poor in quality and also in terms of integration time and field rotation. Only $\sim1$ hr of data is available with a field rotation of $\sim16^\circ$, a coherence time of $2.6\pm0.8$ ms, and a mean seeing of $1\farcs64\pm0\farcs37$. As we show in Figure \ref{fig:LkCa15_comparison}, with deeper observations including more field rotation, ZIMPOL can potentially detect the signal produced by LkCa15\,b \citep{sallum2015} with a CL of 99.99995\%. However, the higher airmass at the Paranal Observatory and the fact that LkCa15 is a fainter guide star may complicate the redetection of the companion candidate, and therefore exceptional atmospheric conditions are required.

In addition to H$\alpha$ also other spectral features like Pa$\beta$ and Br$\gamma$ lines may indicate ongoing accretion processes onto young objects. As an example, \cite{Daemgen2017} used the absence of those lines in the spectrum of the low-mass companion HD106906\,b to infer its mass accretion rate upper limits ($\dot{M}<4.8\times10^{-10}\,M_J/\text{yr}^{-1}$). Their constraint is stronger than the ones we were able to put with our ZIMPOL H$\alpha$ data. Several other studies also detected hydrogen emission lines like Pa$\beta$ from low-mass companions \citep[e.g.,][]{Seifahrt2007, Bowler2011, Bonnefoy2014}, but unfortunately they did not calculate mass accretion rates. 

\subsection{Comparison with existing models}
Two models for planetary accretion are currently used to explain the accreting phase of planet formation: magnetospheric accretion \citep{zhu2015} and boundary layer accretion \citep{Owen_Menou2016}. During magnetospheric accretion, the magnetic field truncates the CPD and hot ionized hydrogen in the closest regions of the disk falls onto the planet following the magnetic field lines. Recombination on the planet surface then produces H$\alpha$ flux. 
For protoplanets, these models predict H$\alpha$ luminosities at least three orders of magnitudes lower than in CTTS, according to equation 22 in \cite{zhu2015},
$$ L_{H\alpha}=4.7\times10^{-6}L_\odot\left(\frac{R_T}{R_J}\right)^2\left(\frac{v_s}{59\text{km s}^{-1}}\right). $$
This is mainly owing to a one order of magnitude smaller infall velocity $v_s$ and a one order of magnitude smaller truncation radius $R_T$ (squared in the $L_{H\alpha}$ equation) due to weaker magnetic fields than in stars. 
We combined the magnetospheric accretion models \citep{zhu2015} with existing detections  in the infrared and evolutionary models. As an example, we present the case of HD100546\,b. According to models \citep{zhu2015}, the observed $L'$ brightness could be emitted by a CPD with inner radius of $1-4\,R_J$ and $M_p\dot{M}$ of $0.2-2.9\times10^{-6}\,M_J^2\text{ yr}^{-1}$. The mass accretion constraints obtained from H$\alpha$ ZIMPOL data would therefore imply that $M_p \gtrsim 31\,M_J$. This result is in conflict with that obtained by \cite{quanz2015} and the AMES-Cond evolutionary models, since the object $L'$ brightness excludes masses larger than $\sim15\,M_J$. This is the mass expected in the case in which the $L'$ flux is only from photospheric emission. Moreover, a 30 $M_J$ object would have significantly shaped the disk morphology and would have been clearly visible in other bands, such as the $K_s$-band, where \cite{quanz2015} could only put upper limits to the companion brightness.
\\
\cite{Szulagyi2017} found that only a minimal fraction of the hydrogen in CPDs might be thermally ionized if the planet is massive and hot enough. Consequently, the disk does not get truncated and ionized material does not get accreted through magnetospheric accretion along the field lines. 
Then, disk material falls directly onto the planet (boundary layer accretion). 
The same authors showed that material falling from the circumstellar disk onto the CPD and the protoplanet shocks, and eventually produces H$\alpha$ line emission both from the CPD and the planet. The contribution to the H$\alpha$ flux is larger from the CPD than from the planet \citep{Szulagyi2017}. These authors also showed that the majority of the accreted gas, however, remains neutral, especially for planets $<10\,M_J$. Hence, the H$\alpha$ flux can only estimate the ionized gas accretion rate and not the total accreted material. According to their simulations, a 10 $M_J$ planet would be accreting at a rate of $5.7\times10^{-8}\,M_J\text{ yr}^{-1}$, producing $L_{H\alpha}\sim7\times10^{-6}\,L_\odot$. This value is on the same order of the limits our data allow us to put on the H$\alpha$ luminosity from known forming protoplanet candidates. Since considering lower planetary masses enhances the mass accretion rate (see equation \ref{eq:accretion_rate}) and higher masses should be visible in other infrared bands, we conclude that either extinction from disk material plays a major role in the nondetection of the existing candidates, or they are false positives resulting from image post-processing.

The comparison of $L_{H\alpha}$ limits from Table \ref{tab:limits} with Figure 7 from \cite{Mordasini2017} indicates that, assuming completely cold accretion, the observed objects may be low-mass ($0.1-1 M_J$) medium accreters ($\dot{M}\sim10^{-10}-10^{-9} M_\odot/\text{yr}$) or higher mass objects ($1-15 M_J$) showing very little accretion ($\dot{M}<10^{-10.5}M_\odot/\text{yr}$). \cite{Mordasini2017} also suggested another possible reason for some of the nondetections in H$\alpha$. If some of the planets, such as HD100546\,b, have not yet completely detached from the disk, they would be cooler and would not be accreting at high accretion rates. In a later phase, they will possibly be able to open a gap and accrete a large amount of material. \\
Another aspect that we did not consider is the effect of the circumplanetary disk inclination on the flux that is emitted. \cite{zhu2015} considered the disk inclination including a factor $1/\cos(i)$, where $i$ is the CPD inclination. Detailed accretion models should investigate the consequences of a tilted protoplanetary disk on $L_{H\alpha}$.


\section{Conclusions}
\label{sec:conclusions}
Imaging in H$\alpha$ is one of the promising techniques to detect forming planets at very small separations. In this context, the SPHERE/ZIMPOL instrument will play a major role in investigating local accretion signatures in circumstellar disks. An important next step is to redetect the previous discoveries of MagAO of H$\alpha$ emission from LkCa15\,b and PDS 70\,b and to study potential accretion variability.
None of the possible protoplanet candidates discovered in the infrared (HD169142\,b, MWC758\,b, and HD100546\,b and c) could be confirmed in this study searching for accretion signatures, implying several possible scenarios. Their mass accretion rates could be lower than our limits and therefore they are currently not detectable. Other explanations are that protoplanetary accretion shows variability and some of the objects are currently going through a period of quiescence, or that extinction effects from disk material absorb a considerable fraction of the light. The study of NIR line diagnostics might reduce the effects of absorption and allow the detection of accretion processes. Furthermore, it is possible that the observed candidates are disk features that have been enhanced by image post-processing \citep{Follette2017,ligi2017}, or our understanding of accretion processes during the formation of giant planets is not correct and, as an example, the use of the CTTS scaling relation is not correct. In order to investigate this, precise simulations of protoplanetary accretion, as well as of disk intrinsic effects (via full radiative transfer), have to be developed and combined with multiwavelength observations spanning from the optical to the (sub)millimeter.

The estimation of upper limits are of particular importance for the study of accretion variability of protoplanets in the future. Continuing surveys for accreting planets could possibly detect H$\alpha$ signatures and combine these with detection limits provided by this work to investigate variability in the accretion processes.
Finally, we emphasize that although a lot of effort was put into the calculation of mass accretion rate upper limits, those values are model and parameter dependent. The H$\alpha$ flux upper limits are, however, the fundamental quantities that were measured from the data and can be directly compared with future observations.


\begin{acknowledgements}
SPHERE is an instrument designed and built by a consortium consisting of IPAG (Grenoble, France), MPIA (Heidelberg, Germany), LAM (Marseille, France), LESIA (Paris, France), Laboratoire Lagrange (Nice, France), INAF - Osservatorio di Padova (Italy), Observatoire de Gen\'eve (Switzerland), ETH Zurich (Switerland), NOVA (Netherlands), ONERA (France), and ASTRON (Netherlands), in collaboration with ESO. SPHERE also received funding from the European Commission Sixth and Seventh Framework Programmes as part of the Optical Infrared Coordination Network for Astronomy (OPTICON) under grant number RII3-Ct-2004-001566 for FP6 (2004-2008), grant number 226604 for FP7 (2009-2012), and grant number 312430 for FP7 (2013-2016). This work has been carried out within the frame of the National Center for Competence in Research PlanetS supported by the Swiss National Science Foundation. SPQ and HMS acknowledge the financial support of the SNSF. GC and SPQ thank the Swiss National Science Foundation for financial support under grant number 200021\_169131. FMe and GvdP acknowledge fundings from ANR of France under contract number ANR-16-CE31-0013. This research has made use of the SIMBAD database, operated at CDS, Strasbourg, France. This work has made use of data from the European Space Agency (ESA)
mission {\it Gaia} (\url{https://www.cosmos.esa.int/gaia}), processed by
the {\it Gaia} Data Processing and Analysis Consortium (DPAC,
\url{https://www.cosmos.esa.int/web/gaia/dpac/consortium}). Funding
for the DPAC has been provided by national institutions, in particular
the institutions participating in the {\it Gaia} Multilateral Agreement.
The authors thank Arianna Musso-Barcucci for the preliminary analysis on HD142527.
\end{acknowledgements}

%
%

\bibliographystyle{aa}
\bibliography{halpha.bib}

\begin{appendix} 
\section{Influence of the beamsplitter on flux measurements}
\label{App:Beamsplitter}
\begin{table}[b!]
\caption{\label{tab:reflectivity_beamsplitter} Resulting signal flux and FPF for different beamsplitter behaviors.}
\centering
\begin{tabular}{lll}
\hline\hline\noalign{\smallskip}
Target                  & Signal flux                                   & FPF \\
 & ($\times1000$, arb. unit) & ($\times10^{12}$) \\\hline
\noalign{\smallskip}
Cnt flux 5\% decreased  & $5.45\pm0.06$    & $9.97\pm4.72$  \\
\noalign{\smallskip}
Cnt flux not changed    & $3.62\pm0.02$    & $1.69\pm1.47$  \\
\noalign{\smallskip}
Cnt flux 5\% enhanced    & $2.84\pm0.03$    & $11.17\pm2.86$  \\
\noalign{\smallskip}
\hline\hline
\end{tabular}
\end{table}
We investigated the effects of the throughput uncertainties of the two ZIMPOL arms resulting from  instrument polarization effects. It is currently not known how the overall throughput to the individual ZIMPOL arms depends on the telescope and instrument configurations. However, it is easy to estimate the overall effect because the Nasmyth mirror of the VLT introduces an instrument polarization of about 4~\%. This is reduced by the first mirror in SPHERE to about 0 to 3~\%, while the following mirrors in the instrument add further positive or negative polarization contributions of about 2~\%, while polarization cross talks (linear $\rightarrow$ circular polarization) reduce the linear polarization. Thus, it is safe to adopt a maximum error of 5~\% for the relative difference (e.g., $T(H_\alpha) = (1 \pm 0.05) T(cont)$) in throughput between the two channels. We therefore tested the impact of an enhancement/decrease by at most 5\% in the continuum flux, analyzing the consequences on the detection of HD142527B and on the contrast performances of our pipeline. 
The signal flux is measured in an aperture of radius 8.3 mas, and the FPF was calculated as explained in section \ref{sec:setup_performance}. The results averaged over a range of PCs (PC=10,15,20,25,30) on the ASDI-processed B\_Ha dataset are shown in Table \ref{tab:reflectivity_beamsplitter}; a central mask of 21.6 mas was applied. As one may expect, the signal flux shows a strong variation of $20-50\%$ in the ASDI images, which is mainly due to the stronger/weaker subtraction of the continuum. The relative difference in this case is increased from the initial 5\% by the ASDI processing, but it should be noted that together with the signal, the noise level also gets increased/decreased, causing the FPF to be less subject to variations.  Indeed, regarding the FPF values, we argue that depending on the arm 1 to arm 2 transmission the confidence of the detection is lower by a factor of $\sim$10 in both extreme cases, which corresponds approximately to a maximum variation of $\sim$0.1 mag in $\Delta$mag. Therefore, we do not expect this effect to have a large impact on the detection limits estimated in this work. Nevertheless, it is important to keep in mind that the calculation of the mass accretion rate of HD142527\,B does not consider this effect and a more accurate description of the instrument behavior is required to correct for it.

\section{FPF analysis of the HD142527 B dataset}
\label{App_1}
\begin{figure*}[h!]
\centering
\includegraphics[width=\hsize]{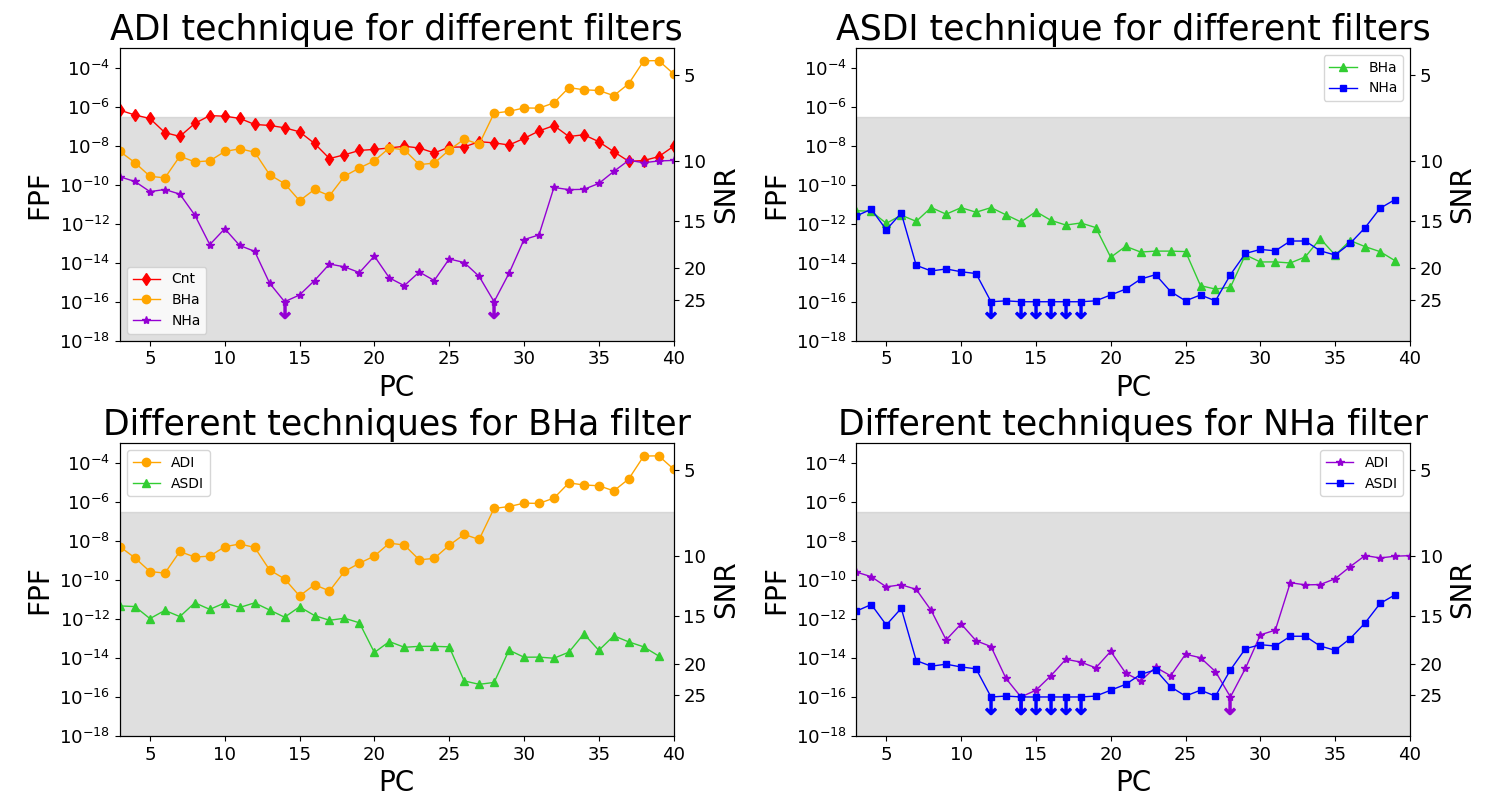}
\caption{Performance comparison for H$\alpha$ imaging with SPHERE/ZIMPOL using different filters (narrow and broad H$\alpha$) and reduction techniques (ADI and ASDI) based on the HD142527 dataset. In all panels the FPF obtained for HD142527 B is shown, as a function of the number of subtracted PCs used in {\tt PynPoint}. 
On the right side of each panel we give the scale of the S/N to improve  understanding of the plot and to compare different instrumental setups. We note again that this does not correspond to the classical $\sigma$ notation.
The gray regions indicates a confidence level for the detection of HD142527 B of at least 99.99995\%, i.e., $>$5$\sigma$ in case of Gaussian noise. Because of the applied corrections for small sample statistics, the border of the gray area does not correspond to an S/N of 5.} 
\label{fpf_HD142527}
\end{figure*}
In order to identify the best strategy for future SPHERE/ZIMPOL observations, we compared the FPF calculated after the subtraction of different numbers of PCs using different techniques and datasets. For the ADI technique we considered three datasets: B\_Ha, N\_Ha, and Cnt\_Ha, of which the last set contained all the images taken with the continuum filter. For SDI and ASDI, we considered the subtraction of the Cnt\_Ha images from the respective H$\alpha$ filter images. All the images had a size of $1\farcs08\times1\farcs08$. For each case, we applied an inner mask of varying size (10.8 mas, 21.6 mas, 32.4 mas) and chose the smallest FPF value as representative value for the detection. The FPF calculation (see Section \ref{sec:setup_performance}) followed the prescription suggested in \cite{mawet2014}. Because of the strong negative wings of the companion in the PSF subtracted images, we decided not to consider the two background apertures closest to the signal as they are not representative of the background and speckle noise.

In the four panels of Figure \ref{fpf_HD142527} we analyzed the FPFs of HD142527 B, obtained using different combinations of techniques and datasets. In the top panels we compare the detection from different filters using the same technique: ADI on the left and ASDI on the right. For the ADI analysis, the B\_Ha and Cnt\_Ha datasets show similar values with a stronger detection in B\_Ha for fewer subtracted components, while the FPF values obtained with the N\_Ha filter are, for a wide range of PCs (from 11 to 32), $\sim$ 5 orders of magnitude lower. The detections with the ASDI technique show a similar trend; there is a stronger detection in N\_Ha, particularly between 10 and 27 PCs. The normal SDI technique, which is not presented in the image, was not efficient enough to properly subtract the stellar PSF and did not reveal the companion. This is probably for two reasons: (1) the central star is actively accreting material and emitting strong H$\alpha$ flux, which cannot be subtracted accurately with the Cnt\_Ha images, impeding the detection of the companion, and (2) PSF shapes are slightly different for different filters due to nonmatching bandpasses.

In the lower panels, we consider the results from the B\_Ha (left) and N\_Ha (right) datasets for ADI and ASDI. In both cases ASDI seems to be more efficient in detecting signals. A larger gain is obtained for the B\_Ha filter, while FPFs obtained with the N\_Ha filter have more similar values, probably due to the minor impact of the continuum subtraction on images taken with the narrow filter with respect to the broad filter. We conclude that the best observing strategy to look for accreting companions in the contrast limited regime with SPHERE/ZIMPOL is to take images in the N\_Ha filter and Cnt\_Ha filter simultaneously and to perform ASDI. 
It is of particular interest that in the case of HD142527, ASDI also performs better than ADI. Indeed, we could expect that the presence of a clear signal in the continuum would have strongly compromised the detection with ASDI. On the contrary, the detection is even stronger, implying that the subtraction of the stellar pattern is much more important than the self-subtraction of the companion, boosting its S/N.
We note, however, that observing fainter objects might cause the data to be readout noise limited. In this case, the B\_Ha filter might be preferred to the N\_Ha filter. This hypothesis, however, should be confirmed with a fainter source than the bright M-dwarf HD142527B.

\section{Impact of field rotation and total integration time}
\label{App_2}
\begin{figure}[h!]
\centering
\includegraphics[width=\hsize]{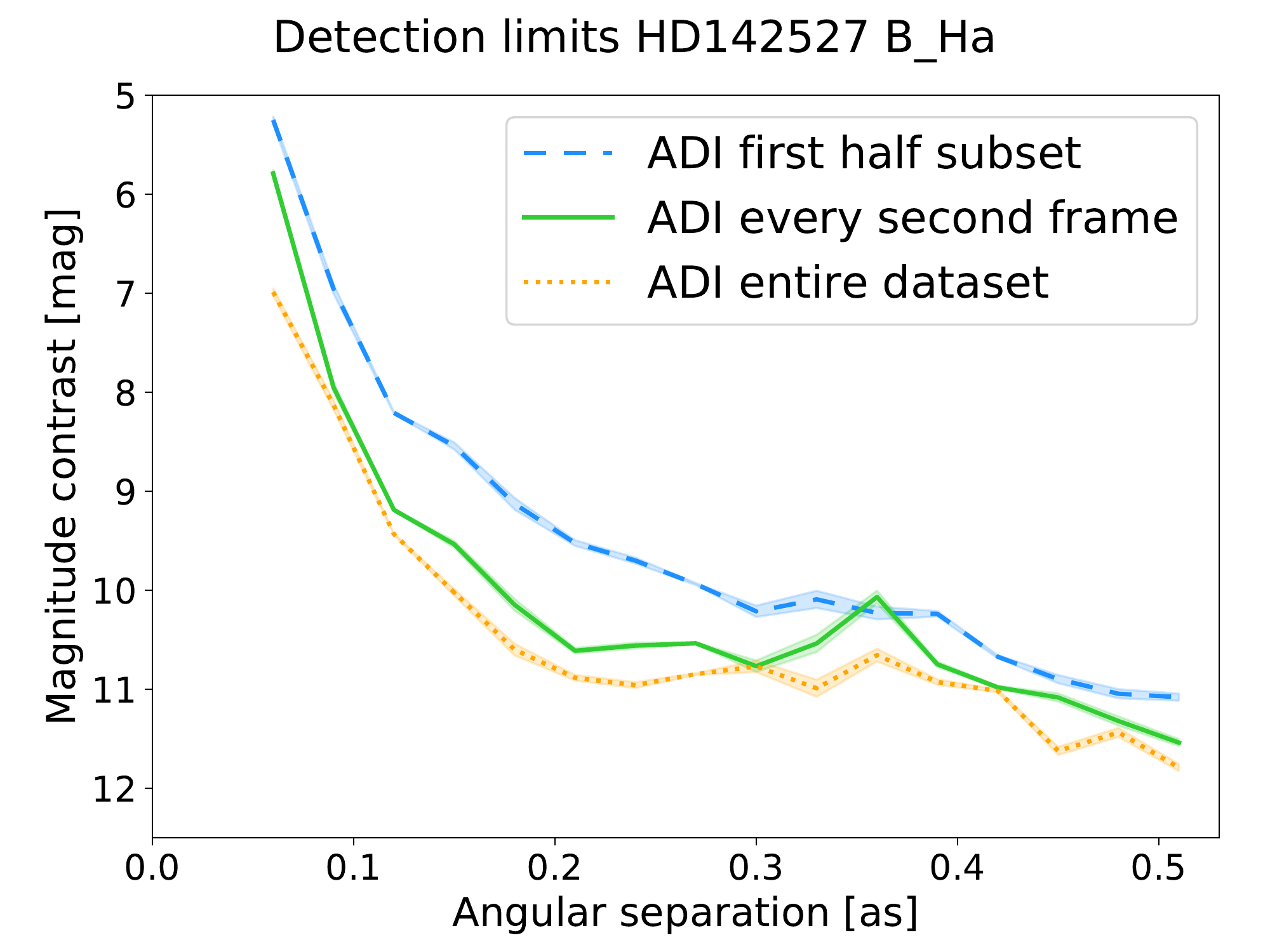}
\caption{Contrast curves calculated for the ``first'' (blue dashed line) and the ``every second frame'' subsets (green solid line), and for the entire B\_Ha dataset of HD142527 (orange dotted line). }
\label{fig:obs_param}
\end{figure}

In addition to the best instrumental setup for the detection of accreting objects in H$\alpha$ imaging, we also investigated the effect of two observational parameters on the achieved upper limits: the field rotation and integration time on target. Two subsets were created from the HD142527 B\_Ha data. The first was composed of every second frame of the dataset, while the second only included the first half of the dataset frames. In this way, the field rotation of the first subset is twice that of the second subset, while the integration time is the same for the two subsets. Figure \ref{fig:obs_param} shows the resulting contrast curves, calculated in the same way as described in Section \ref{sec:setup_performance}. The dashed blue line represents the subset composed of the first half of the dataset, which allows us to reach $\sim9.4$ mag of contrast at $0\farcs2$. The green solid line shows the contrast limits estimated from the subset composed of every second frame. It is clear that at all separations, this subset allows us to detect fainter objects than the other subset and at $0\farcs2$ the difference reaches 1.1 mag. Finally, the entire dataset allows us to go, at the same distance, another 0.3 mag deeper. At least for this dataset, the field rotation seems to play a very important role, allowing a better modeling and subtraction of the stellar PSF.

\section{Is a companion candidate orbiting HD135344B?}
\label{app:HD135344B_companion}
\begin{figure}[t!]
\centering
\includegraphics[width=0.8\hsize]{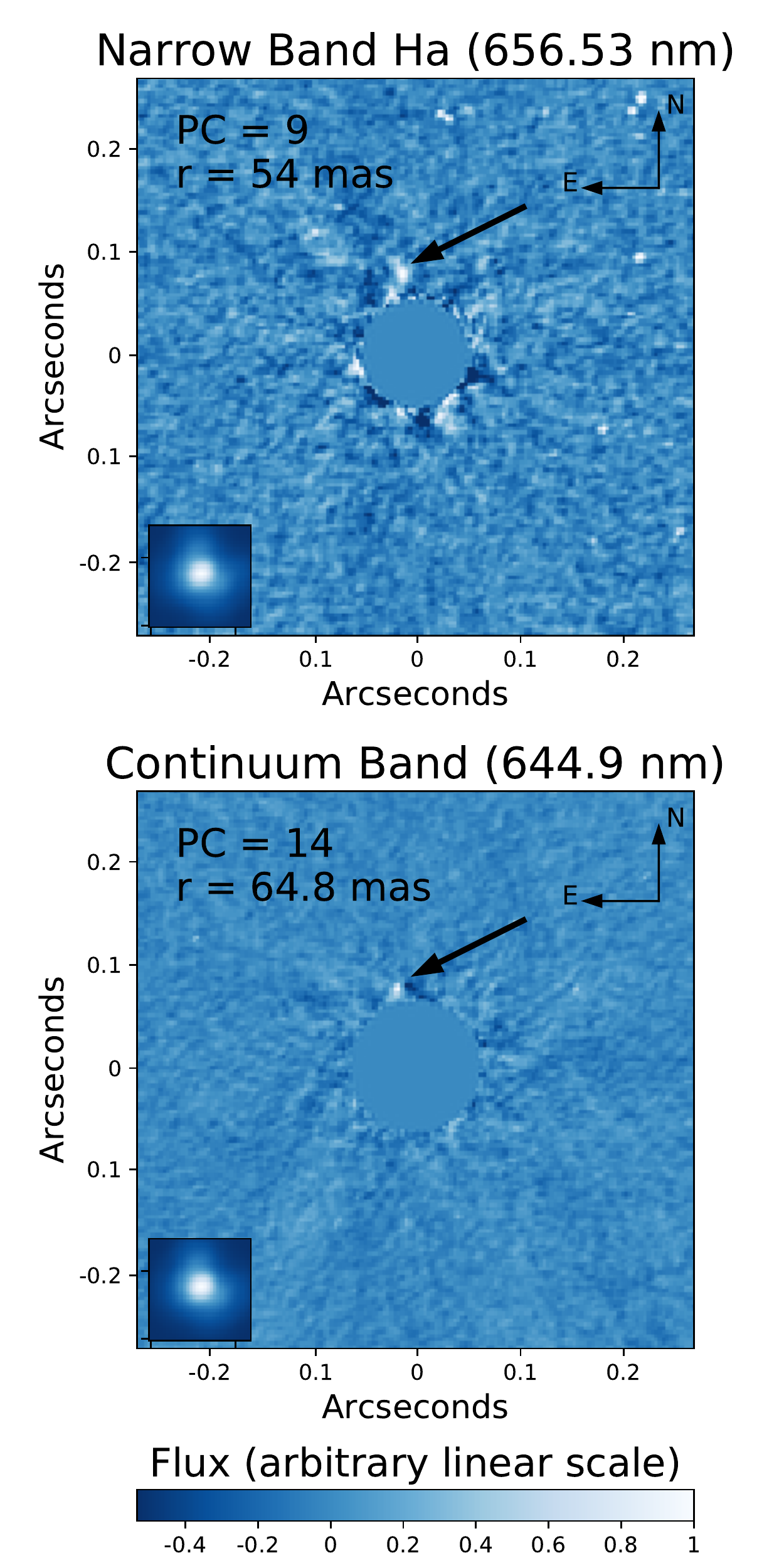}
\caption{Lowest FPF images of HD135344 B (top panel: N\_Ha filter; bottom panel: Cnt\_Ha filter). The radius $r$ of the inner mask and the number of subtracted PCs are given in each panel. The location of the tentative companion candidate is indicated by the arrow (see Section 4.2.1.).}
\label{fig:HD135344Bb_image}
\end{figure}

We visually inspected the final PSF-subtracted ADI images of HD135344 B, which showed a potential signal north of the star. The feature is persistent in the N\_Ha and Cnt\_Ha datasets for different mask radii (e.g., $0\farcs02$, $0\farcs03$, $0\farcs04$, $0\farcs05$, and $0\farcs06$) and over a wide range of PCs (6-21). In particular, when using larger mask radii, the close-in speckles are removed and the signal appears to be stronger. We then investigated smaller images (101$\times$101 pixels) with the same technique and confirmed the signal for different mask radii and PCs. Next, we examined the ASDI images and found that the signal is present once again in different reductions, but appears fainter. If the signal is from a physical source, this is expected from an accreting object emitting H$\alpha$ line radiation. This signal is shown in Figure~\ref{fig:HD135344Bb_image} in the N\_Ha and in the Cnt\_Ha filter for the parameter setups that yield the lowest FPFs, which are $5.9\times10^{-5}$ for N\_Ha and 0.0015 for Cnt\_Ha.
A careful look at the bright signal in the Cnt\_Ha images raises doubts on the nature of its source as it is very compact and does not have a PSF-like shape. 
Furthermore, the signal has high FPFs, with a minimal value of $\sim0.0015$, which is not statistically significant enough to claim a detection.

The signal in the N\_Ha frames has a morphology resembling that of a faint physical source. Varying the number of PCs seems to influence the apparent shape and location of the signal, as expected from faint close-in objects with low S/N when subtracting the stellar PSF. Even though for 9 PCs the FPF reaches a minimal value of $\sim5.9\times10^{-5}$, the FPFs for 7-17 PCs are in the range $10^{-3}-10^{-2}$, which does not give us sufficient confidence to claim a detection.

As a final check, we used the Hessian matrix approach as described in Section \ref{sec:characterization} to perform a signal characterization. We ran the algorithm for PCs between 7 and 17 (where the final images showed a clear signal) with a central mask of radius 57.6 mas and a ROI of $8\times8$ pixels. The other parameters were kept identical to the analysis performed on HD142527 B. The signal appears to be located at a separation of $71.1^{+4.8}_{-4.2}$ mas with a PA of $(19.1^{+2.2}_{-2.8})^\circ$. The contrast was measured to be $8.1\pm0.4$ mag. As visible in the error bars, the positions found are spread over a range of $\sim9$ mas, which corresponds to $\sim2.5$ pixels. Normally, a physical point source should be less affected by systematics introduced by the PSF subtraction process. However, a low S/N object at a separation of 71.1 mas is more difficult to measure properly and a larger spread in the recovered positions could be the result. A similar note can be made for the contrast values, which span over a range of $\sim0.8$ mag. We conclude that to settle this issue and fully understand the origin of the signal in the  H$\alpha$ filter, a dataset with higher S/N would be required. 

\section{Frame selection for the HD100546 dataset}
\label{App_3}
\begin{figure}[h!]
\centering
\includegraphics[width=\hsize]{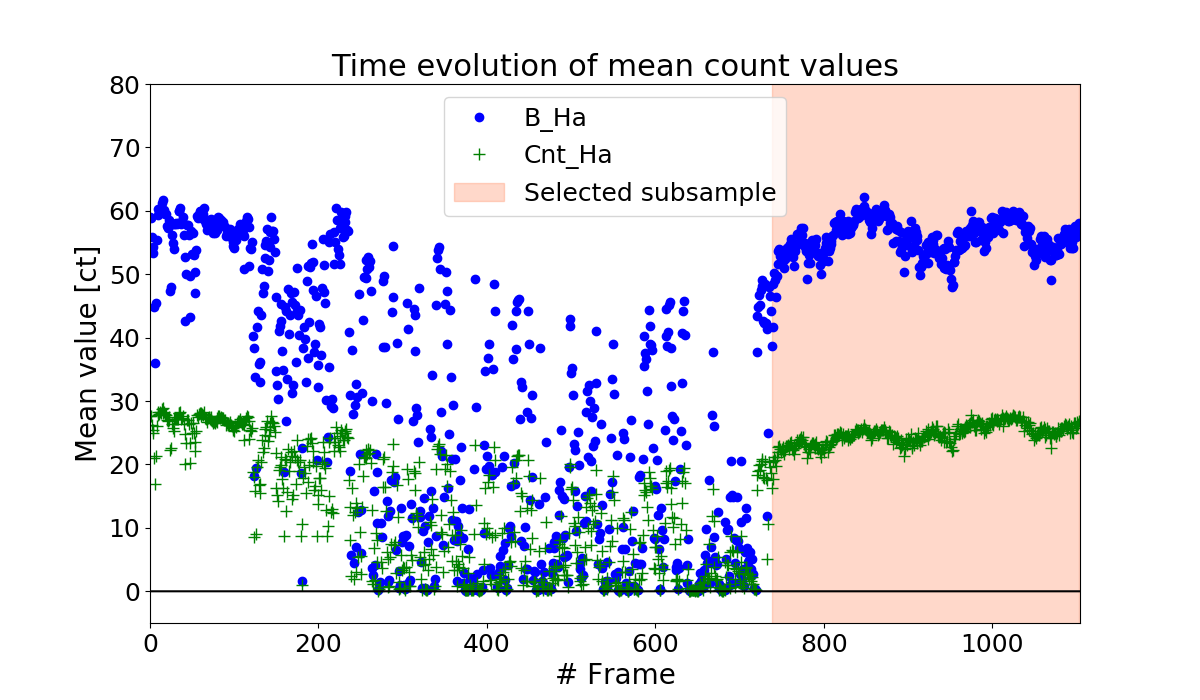}
\caption{B\_Ha (blue circles) and Cnt\_Ha (green crosses) mean count rates as a function of the image number in the observing sequence. The shaded region at the end represents the subset of frames that was chosen for the analysis. }
\label{fig:HD100546_frame_sel}
\end{figure}

As briefly described in Section \ref{sec:Analysis_HD100546}, the large HD100546 dataset (1104 frames, cf. Table \ref{tab_obs}) was taken in unstable conditions, which made a frame selection necessary. To determine a frame selection metric, we plotted the mean count value per image (image dimensions $1\farcs08\times1\farcs08$ pixels, see Figure \ref{fig:HD100546_frame_sel}).
It turned out that three phases could be identified within the observing run: a short initial phase of stability with some outliers (120 frames), a long period of 619 frames where the mean count values spanned a range between $\sim0$ and $\sim55$ counts per pixels, and, finally, a large amount of stable frames at the end of the observations (mean pixel value $\sim55$ in B\_Ha). We decided only to  keep the images of the last stable period, composed of the frames $739-1104$ to perform our analysis. This subsample has a total on-target integration time of 61 minutes and its field rotation is $\sim20.7^\circ$.

\end{appendix}

\end{document}